\documentclass[aps,prx,onecolumn,notitlepage,10pt,amsmath,amssymb,superscriptaddress]{revtex4-2}

\usepackage{xcolor}
\usepackage{graphicx}
\usepackage[version=4]{mhchem}
\usepackage{microtype}
\usepackage{tikz}
\usetikzlibrary{arrows.meta,positioning,shapes,fit,backgrounds,calc}
\usetikzlibrary{arrows,decorations.pathreplacing,calligraphy}
\usepackage{xurl}
\usepackage[bookmarks=false,colorlinks]{hyperref}
\hypersetup{linkcolor=purple,citecolor=red,filecolor=purple,urlcolor=green}
\usepackage{booktabs}
\usepackage{makecell}
\usepackage{placeins}

\begin{document}
\raggedbottom
\title{A Structural Map of Potential Correlated Electron Molecular Orbital Materials}
\author{Md. Rajbanul Akhond}
\email{mdakhond@iu.edu}
\author{Alexandru B. Georgescu}
\email{georgesc@iu.edu}
\affiliation{%
Department of Chemistry, 800 East Kirkwood Avenue, Indiana University, Bloomington, Indiana 47405, United States}
\date{\today}

\begin{abstract}
  Correlated electron molecular orbital (CEMO) materials host emergent electronic states built from molecular orbitals localized over clusters of transition metal ions in a solid, yet have historically been discovered sporadically and generally been treated as isolated case studies. Here, we present criteria and tools to classify and analyze these materials, provide a database of identified materials with cluster-specific symmetry analysis, including cluster point groups, effective cluster sublattice dimensionality. Our electronic structure analysis shows that narrow bands in cluster materials can arise either from relatively isolated cluster molecular orbitals or from a mixed mechanism in which cluster localization coexists with frustrated-lattice flat-band topology. We discuss how CEMO properties do not arise from metal clusters in a solid alone: rather they depend on electron count, symmetry, frustration and correlation strength. By integrating cluster classification with  flatband lattice topology and electrochemistry-relevant information, we provide further relevant information to multiple scientific communities.
  Applying this approach in a high-throughput screen of 34,548 compounds yields potential 2,627 stable or metastable CEMO compounds with isolated cluster motifs. The accompanying open dataset, Cluster Finder software, and interactive web platform enable systematic exploration of cluster-containing electronic solid state materials and establish a starting point for discovering a new class of correlated quantum materials with a wide array of properties.

\end{abstract}
\maketitle

\section{Introduction}

Correlated electron molecular orbital materials (CEMO) - or correlated electron quantum cluster materials - are solid state, inorganic materials where emergent correlated electron states are obtained from molecular orbitals extended on clusters of transition metal ions, as opposed to atomic d- and f-orbitals. These have also been refered to as 'orbital molecules' in solids \cite{attfield_orbital_2015}. Extended molecular orbitals in atomic clusters within solid-state inorganic materials resemble molecular orbitals in organic and inorganic chemistry, forming extended states across multiple ions. One important consequence of cluster molecular-orbital formation is the emergence of narrow or flat electronic bands near the Fermi level. Such bands enhance electron-electron interactions and can support Mott insulating states, unconventional magnetism, spin-liquid-like behavior, and superconductivity \cite{grytsiuk_nb3cl8_2024,gall_synthesis_2013,mourigal_molecular_magnet_2014, marini_superconducting__2021}. In contrast to flat bands that arise purely from destructive-interference mechanisms in frustrated lattices, the flat or weakly dispersive bands in CEMO materials can originate from spatial localization within finite transition-metal clusters. As the clusters host electrons in molecular-like orbitals, they offer platforms to investigate novel quantum states and phenomena arising from electron correlations and spatial confinement. Their unusual geometry allows for a variety of applications across chemistry, physics, energy, electronics, and quantum information science \cite{kelly_nonpolar--polar_2019}. 

In molecular chemistry, molecular orbitals emerge from the interaction between atomic orbitals, creating bonding and antibonding states that govern stability, and the resulting electronic properties. In cluster materials, the symmetry and electron filling of the molecular orbital states plays a key role in cluster formation and materials' stability, as does 'just right' electron correlation strength \cite{kumari_molecular_2025}; strong electron correlations may break the molecular orbital states and lead to local d-orbital states (charge order), while delocalized orbitals may lead to longer-range metal-metal bonding motifs (e.g. linear chains). The materials' unique geometry can lead to novel effects. For example, 2D trimer material \ce{Nb3Br8}, leads to spin $S=1/2$ states on \ce{Nb3} triangles (making it a potential spin-liquid candidate material), while 2D polar modes perpendicular to the plane have led to these materials being used as the first field-free Josephson diode for quantum computing applications \cite{wu_field-free_2022}. \ce{Mo3O8} based cluster magnets \cite{nikolaev_quantum_2021}, including \ce{Li2InMo3O8} show unconventional magnetic ordering \cite{gall_synthesis_2013} while, \ce{LiZn2Mo3O8} forms collective molecular magnets \cite{mourigal_molecular_magnet_2014} and \ce{Li2ScMo3O8} has been studied as a potential spin-liquid \cite{haraguchi_spin--liquid_2015}, depending on the nature of interaction and disorder. Another promising material from this class is $Li_xScMo_3O_8$ - which can be doped electrochemically, and as a result, the magnetic state can be tuned. \cite{wyckoff_electrochemical_2023} Another family of materials with Mo clusters are the Chevrel phases, which contain $Mo_{3n}$ clusters, of which the most well known have the general formula $M_xMo_6X_8$, where X is a chalcogen like S or Se, and M is an intercalated metal. These phases not only show promise as electrode materials for monovalent and multivalent ion batteries \cite{mei_chevrel_2017} but also as possible superconducting phases\cite{banerjee_evidence__2024, marini_superconducting__2021} with some even showing reentrant superconductivity \cite{goslawska_reentrant__1998}, where a low temperature magnetic state removes the superconducting state at T=0K, which is present above the transition temperature.

\begin{figure}
  \centering
  \includegraphics[width=1\linewidth]{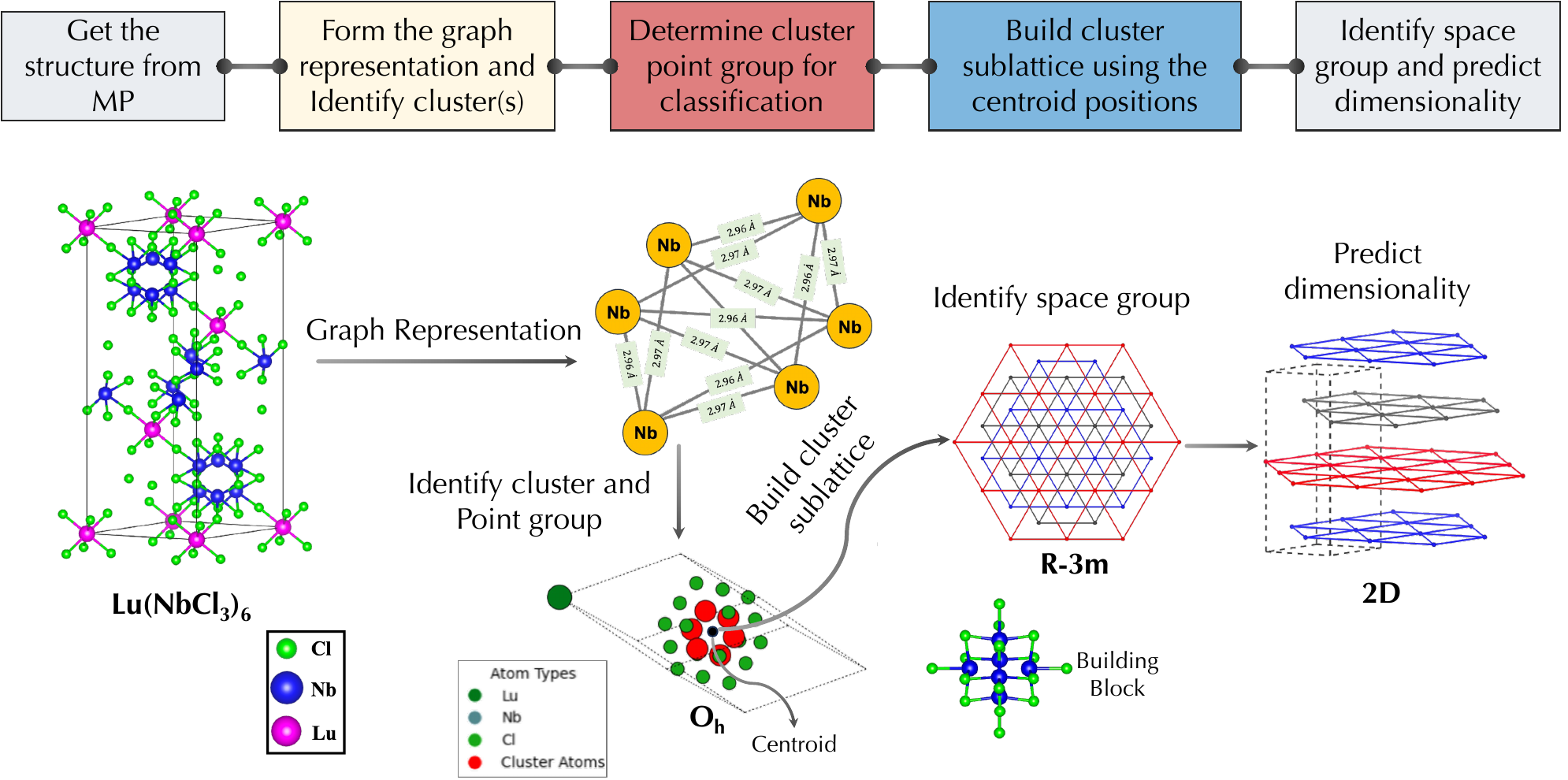}
  \caption{Flow chart of how the Cluster Finder identifies, classifies and determines different geometrical properties of the clusters in materials. }
  \label{fig:flowchart}
\end{figure}

Another material family is that of lacunar spinels, which form tetrahedral tetramers. They  are a family of compounds with the general formula \ce{GaM4X8}, where M represents transition metals and X denotes chalcogens. These materials may display relevant properties, including Jahn-Teller induced ferroelectricity \cite{xu_unusual_2015}, multiferroicity \cite{singh_orbital-ordering-driven__2014, ruff_multiferroicity_2015}, large negative magnetoresistance \cite{dorolti_half-metallic_2010}; their molecular orbital-derived bands can be measured optically \cite{opticalspinel}.

\begin{figure}
  \centering
  \includegraphics[width=1\linewidth]{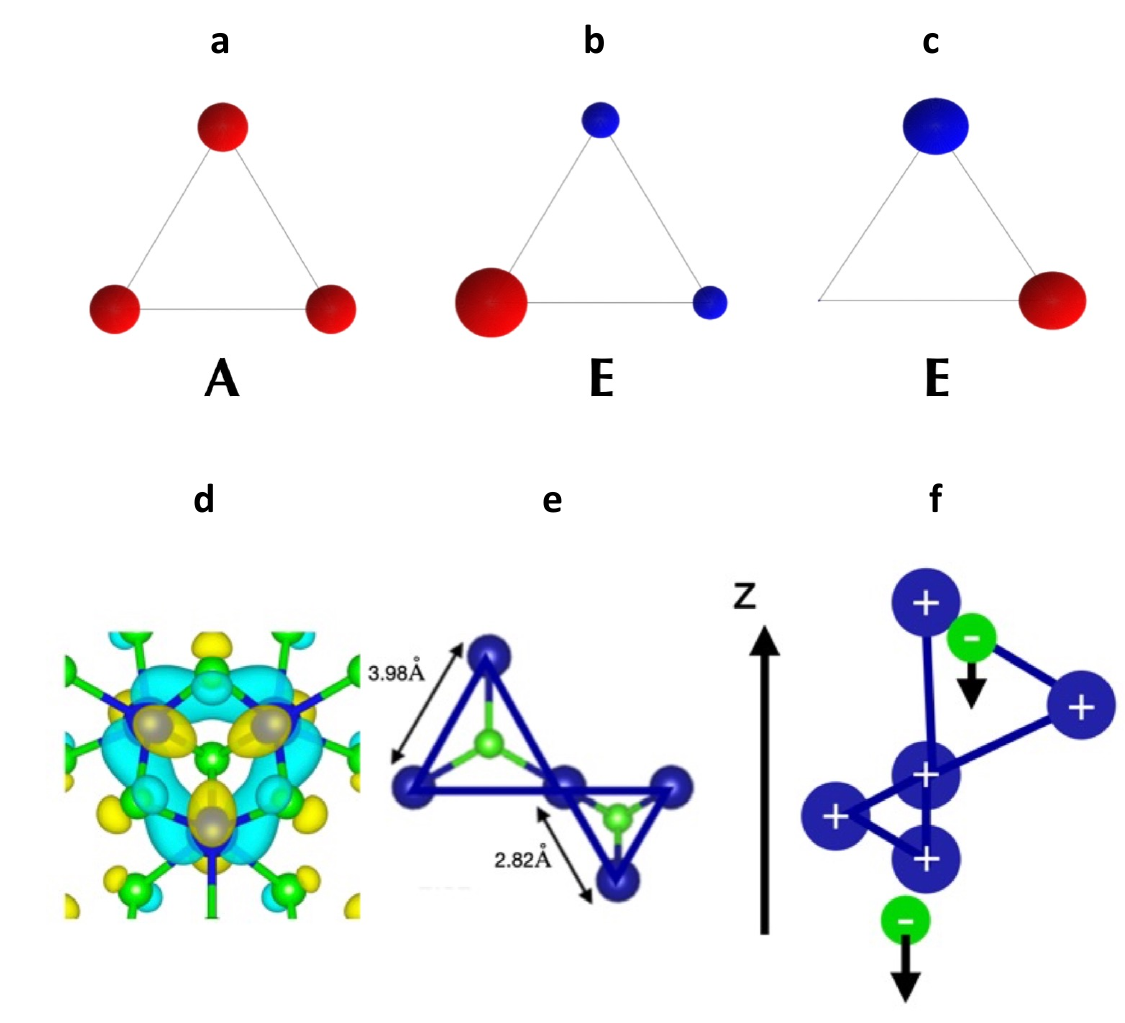}
\caption{a,b,c Simple eigenbasis for a particle on a triangle (trimer) model system using s-like isomorphic orbital basis, showing the relative phases of the s-orbitals forming the molecular orbitals for the bonding A Molecular orbital, and two degenerate T Molecular orbitals. d: DFT results showing a Molecular Orbital on a $Nb_3$ clusters in triangular trimer material \ce{Nb3Cl8}, as obtained by plotting an isosurface of a wavefunction at the $\Gamma$ point (Nb: blue, Cl: green). e) This behavior is associated with a breathing distortion that stabilizes triangular motifs leading to an f) polar structural mode\cite{kelly_nonpolar--polar_2019}.}
\label{fig:orbital}
\end{figure}

Moreover, this molecular orbital (MO) picture is required to understand correlated electron materials with transition metal clusters, as exemplified in Nb$_3$X$_8$ (X=Cl, Br, I) \cite{kumari_molecular_2025, grytsiuk_nb3cl8_2024, hugomalte}, Na$_2$Ti$_3$Cl$_8$ \cite{kelly_nonpolar--polar_2019}, and  GaV$_4$S$_8$ \cite{kim_molecular_2020}, where it builds upon and moves beyond atomic descriptions by capturing intracluster hybridization and correlations that drive cluster stability, a novel polar mode-induced multiferroicity, and Mott gap opening respectively.

For cluster stability and formation, the MO picture reveals that optimal electron filling (e.g., 6-8 electrons per trimer in Nb$_3$X$_8$) driven by symmetry (occupying bonding MOs while avoiding antibonding ones), with intermediate correlation strength balancing delocalization and localization to prevent over-clustering or charge order are needed \cite{kumari_molecular_2025}.

MO-driven distortions can activate polar modes \cite{kelly_nonpolar--polar_2019}, leading to coupled electric polarization and magnetism, where noncentrosymmetric phases emerge without traditional pseudo-Jahn-Teller d$^0$ \cite{bersuker2013pseudo} or lone-pair mechanisms, enabling magnetoelectric effects in kagome layers (Figure \ref{fig:orbital}). Finally, the MO picture explains Mott insulation through narrow, flat bands from cluster-localized states, where single-site methods fail but cluster MO-DMFT naturally opens gaps \cite{hugomalte}, and leading to potential spin-liquid phases. Another similar material from these Lacunar spinel phases is \ce{GaTa4Se8} which may display topological superconductivity induced by pressure \cite{park_pressure-induced_2020}.

Despite their wide range of properties, these materials have generally been discovered sporadically. In this paper, we will discuss how to systematically discover, indentify and classify materials with transition metal clusters which may display correlated electron properties resulting from molecular orbital formation, as well as related materials with strong metal-metal bonding. The resulting dataset alongside with the "Cluster Finder" python library and the "Cluster Explorer" web application offer platforms to analize and explore materials displaying this type of structural and electronic pattern. In addition, we add easy-to-use tools to visualize their potential flatband topology, and electrochemical properties as obtained from high throughput databases, as discussed below.

\section{Cluster Motifs}

\begin{figure}[tbp]
\centering
\includegraphics[width=0.75\linewidth]{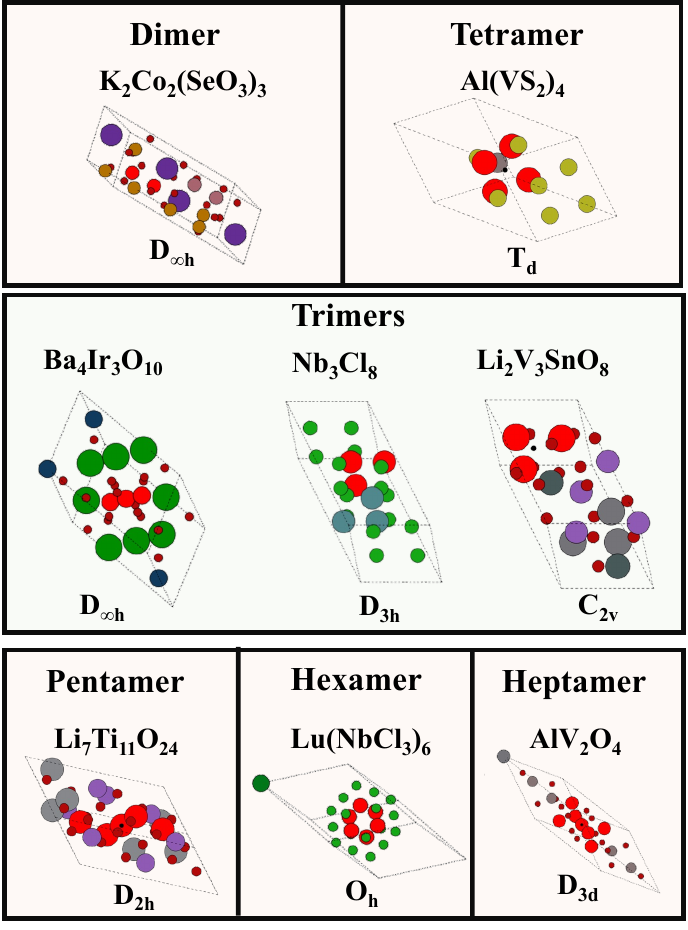}
\caption{Different type of cluster motifs discovered by the Cluster Finder code with cluster point group classification.}
\label{fig:motifs}
\end{figure}

Attfield \cite{attfield_orbital_2015} gave a catalog of materials which contains these extended electronic orbital states, and labeled MOs as 'orbital molecules', while Khomskii \cite{khomskii_orbital_2021} later expanded them as 'Molecules in Solids'. However, the space of materials with cluster motifs which can sustain these extended MOs or orbital molecules is vast and underexplored. In this work we use the CEMO term to avoid confusion with molecular/non-periodic cluster systems, but still reference their electronic building blocks as molecular orbitals using standard chemistry terminology. Using our Cluster Finder code we have found a diverse collection of transition metal compounds with cluster motifs, which can be categorized by size and by geometry (Table \ref{pgt}). For example, clusters of size three (trimers) can be present in perfectly equilateral triangles ($D_{3h}$), isosceles triangles ($C_{2v}$) or linear trimers ($D_{\infty h}$)(Figure \ref{fig:motifs}). Finally, these cluster motifs are spatially distributed according to a diverse set of translational and space group symmetries. The arrangement of the clusters in the solid can reduce the effective dimensionality of these compounds (e.g., Ir  trimers in \ce{Ba4Ir3O10} show effective 1D chain behavior. Nb triangular trimers in \ce{Nb3Cl8} are arranged in a 2D triangular lattice Figure \ref{fig:dim}).

To validate the generality of the algorithm, we applied Cluster Finder to a chemically and structurally diverse set of known cluster materials spanning dimers, linear trimers, triangular trimers, tetramers, pentamers, hexamers, and heptamers. In each case, the method recovers the expected cluster size and point-group classification, demonstrating that the graph/splitting procedure is robust and not limited to a single structural family. For example, we found that it successfully identifies known quantum cluster materials including dimer material \ce{K2Co2(SeO3)3} \cite{zhong_frustrated_2020}, linear trimers in \ce{Ba4Ir3O10} \cite{cao_quantum_2020}, tetramers in Mott insulating \ce{Al(VS2)4} \cite{kim_molecular_2020} among many others (Figure \ref{fig:motifs}). More importantly, we have uncovered several previously unrecognized candidate materials. One notable example is \ce{Lu(NbCl3)6}, where we identify a \ce{Nb6} cluster with an intra-cluster Nb-Nb distance of \(2.96\,\text{\AA}\). Cluster symmetry analysis reveals an octahedral (\(O_h\)) point group embedded within an \(R\bar{3}m\) rhombohedral lattice, forming two-dimensional layers with triangular packing (Figure~\ref{fig:ucluster}). This compound has been experimentally synthesized \cite{ihmaine_structure__1988}, though its cluster-based electronic properties have not yet been explored. All the cluster motifs we have discussed so far are mainly isolated. There are other types of pseduo-cluster connectivity we have identified namely, shared and extended. Details for how these have been identified and classified can be found in the Supporting Information.

To make the materials we have identified available and easy to visualize, as well as their cluster motifs, we have built an online interface, "Cluster Materials Explorer" (Figure \ref{fig:cme}). This includes information on the cluster point group, metal-metal bond lengths, effective dimensionality, as well as symmetry-allowed physical properties, including enantiomorphism, optical, polarity, and piezoelectricity (Figure \ref{fig:venn}). We have added possible frustrated flatband and electrochemical information according to the crystal net catalog \cite{neves_crystal_2024} and Materials Project battery dataset \cite{jain_commentary_2013}, respectively. We mention that the flatbands in the crystal net catalog use a different methodology and focuses on flatbands arising from frustration, rather than cluster molecular orbital localization. All the materials can be viewed in an interactive way in the interface and each material is linked to its Materials Project (MP) and/or ICSD structure if available. The methodology used to identify, analyze and classify these materials will be discussed below.

\section{Chemical Origins of Narrow Bands in Transition-Metal Cluster Solids}

\begin{figure}[h!]
\centering
\includegraphics[width=0.8\linewidth]{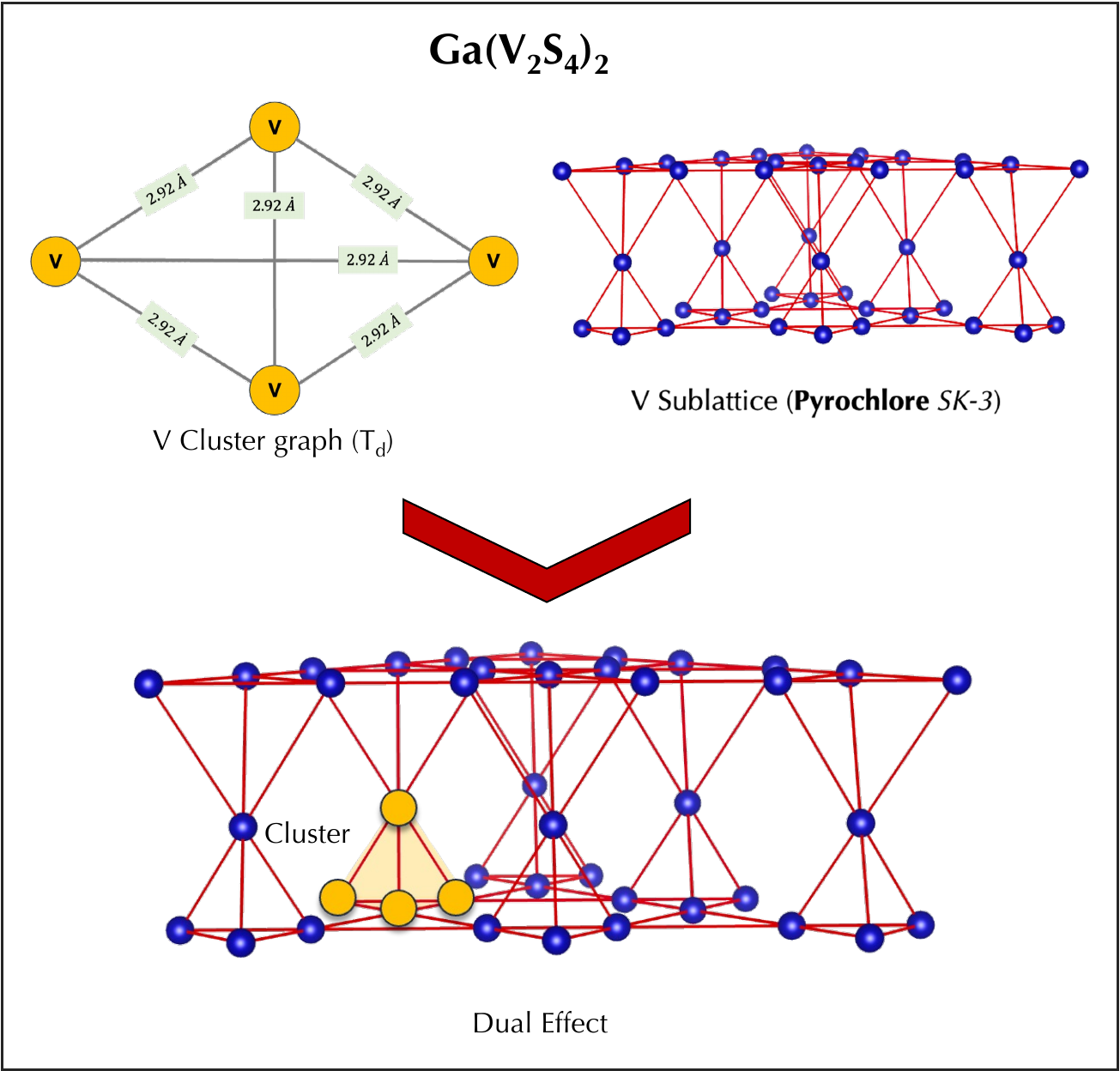}
\caption{Dual effect illustrated for lacunar spinel \ce{Ga(V2S4)2}. Top left: graph representation of the local V$_4$ tetramer cluster with V--V bond lengths. Top right: the V sublattice. Bottom: embedding of the local cluster motif within the extended sublattice, highlighting how cluster molecular-orbital localization coexists with a flat-band-supporting lattice topology.}
\label{fig:graph}
\end{figure}
Narrow bands in transition-metal cluster solids can arise from at least two chemically distinct mechanisms: localization into finite cluster molecular orbitals and destructive-interference flat bands generated by frustrated lattice topology. These mechanisms are often discussed separately, but the cluster materials identified here show that they can appear independently, coexist in the same compound, or fail to produce a clean molecular-orbital manifold despite short metal-metal contacts. Distinguishing these limits is key to identify true correlated electron molecular orbital candidates, because the presence of a structural cluster motif alone does not guarantee cluster-derived low-energy electronic states.

 To understand these mechanisms, we have added lattice frustration induced flat band information to the cluster materials dataset, indentified according to the crystal net tight binding (TB) method \cite{neves_crystal_2024}. This method identifies non-trivial flat bands in materials via a high-throughput workflow: extracting elemental sublattices from the Materials Project database, constructing nearest-neighbor tight-binding models with Hamiltonian \( H = - \sum_{\langle i,j \rangle} t_{ij}(c_i^\dagger c_j + c_j^\dagger c_i) \), eliminating trivial isolated bands, computing bandstructures on a k-point grid to detect flat bands with dispersion \(\leq 0.1\%\), classifying lattices topologically, and filtering for robust candidates based on structural criteria, including atom count and bond distances, yielding thousands of unique flat band nets. Some of the possible stable cluster materials which are also frustrated flatband materials according to the method are shown in Table \ref{fbt}, showing a wide varity of cluster materials containing different number and type of flat band(s). When comparing the cluster dataset with crystal net catalog, there are 312 materials with both CEMO isolated motifs and frustrated lattice. 

\begin{figure}
\centering
\includegraphics[width=0.9\linewidth]{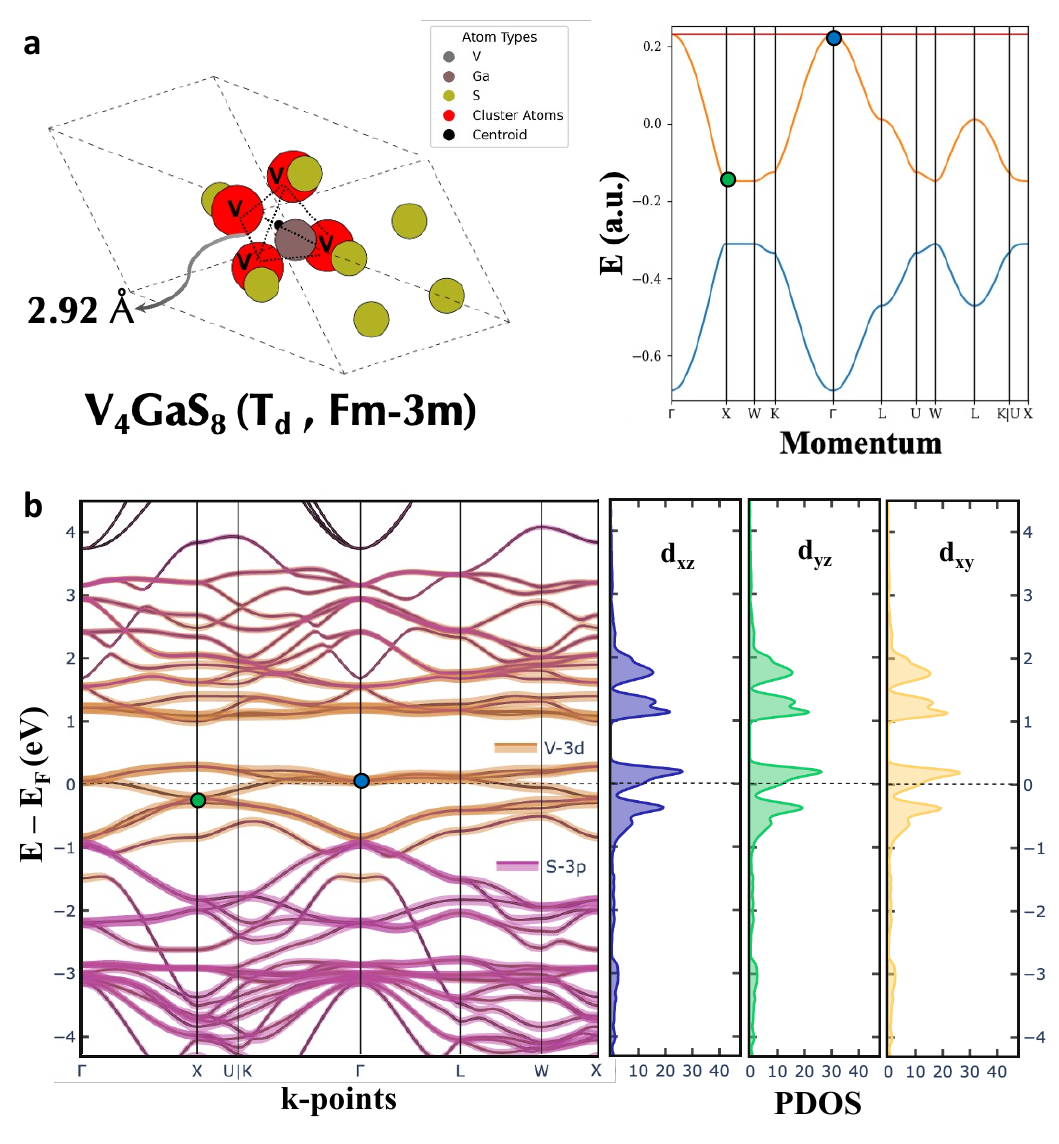}
\caption{Cluster-derived narrow bands in lacunar spinel \ce{GaV4S8}. (a) The \ce{V4} tetramer motif embedded in the frustrated V pyrochlore-like sublattice and the corresponding crystal-net tight-binding band structure, showing narrow bands near the Fermi level, including a dispersive feature at X (green) and a degeneracy at $\Gamma$ (blue). (b) Similar band topology in the DFT band structure and orbital-projected density of states (PDOS), showing that the states in this energy window are dominated by the V-$3d$ $T_2$ manifold ($d_{xz}$, $d_{yz}$, and $d_{xy}$), as expected for the tetrahedral crystal field of the $T_d$ \ce{V4} cluster.}
\label{fig:dual}
\end{figure}

An interesting outcome of this overlap can be seen in materials with breathing distortions. \ce{Nb3Cl8} is a 2D material with a kagome lattice that shows breathing distortion. Without a breathing distortion, the triangles formed by Nb are identical in size, there is no cluster formation, and there is no flat band near the Fermi level (Figure \ref{fig:bd}a) due to the strong role of the Cl ions in suppressing the kagome flatband formation. However, with a breathing distortion, which is the stable structure for \ce{Nb3Cl8}, trimer clusters form and an associated flat band appears near the Fermi level (Figure \ref{fig:bd}b) signifying that these flat bands are comparatively robust with respect to perturbation from the chemical environment. An example of a stable kagome lattice without a breathing distortion is \ce{Co3(SnS)2}. This material should have a flat band due to lattice frustration (Figure \ref{fig:real}a); however, DFT results show the absence of such an isolated flat band (Figure \ref{fig:real}b). Another related but distinct example can be found in lacunar spinel \ce{GaV4S8}, where breathing distortion in a 3D pyrochlore-like lattice partitions the V network into well-defined \ce{V4} tetramers while preserving the frustrated connectivity of the underlying sublattice thus, embedding the clusters in a sublattice of the same material (Figure \ref{fig:graph}). In Figure \ref{fig:dual}a, the crystal-net TB result shows narrow bands near the Fermi level. In Figure \ref{fig:dual}b, the DFT band structure and PDOS show that this low-energy manifold with frustrated band topology survives in the real material and is dominated by V-$3d$ states of cluster character (the $T_2$ manifold). Therefore, \ce{GaV4S8} provides a direct example in which breathing distortion creates cluster molecular orbitals inside a lattice that also carries flat-band topology, giving rise to narrow bands with a dual origin. 

\begin{figure}
\centering
\includegraphics[width=0.88\linewidth]{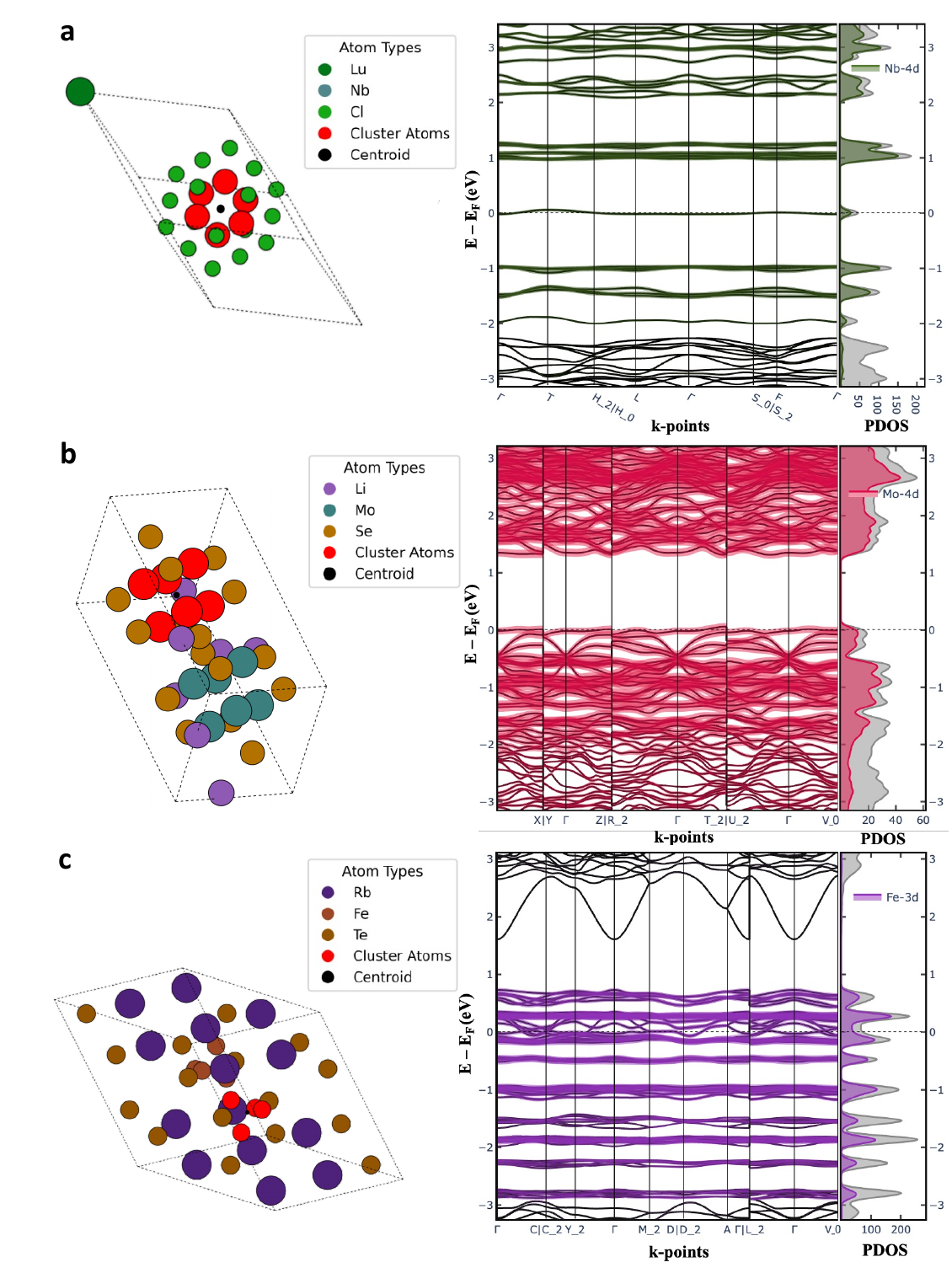}
\caption{DFT band structures and projected densities of states (PDOS) for representative cluster materials. Left: crystal structures with the transition-metal clusters highlighted. Right: band structures and PDOS. (a) \ce{Lu(NbCl3)6}, containing an isolated \ce{Nb6} hexamer cluster, with narrow Nb-$4d$ bands near the Fermi level. (b) \ce{Li7(Mo3Se4)4}, containing \ce{Mo6} hexamer, where the low-energy electronic structure is dominated by Mo-$4d$ states with dual character. (c) \ce{Fe16Rb28Te32}, containing Fe-based tetramer \ce{Fe4} clusters, with multiple weakly dispersive Fe-$3d$ bands around the Fermi level.}
\label{fig:dfts}
\end{figure}

Figure \ref{fig:dfts} shows that this dual cluster-and-lattice picture extends beyond known materials to a broader set of unexplored cluster materials. In Figure \ref{fig:dfts}a, the isolated hexamer \ce{Nb6} cluster in \ce{Lu(NbCl3)6} produces a compact manifold of narrow Nb-$4d$ bands near the Fermi level, representing a relatively clean cluster-molecular-orbital limit. Figure \ref{fig:dfts}b shows a more mixed case in \ce{Li7(Mo3Se4)4}, where the \ce{Mo6} cluster still controls the low-energy electronic structure, but with more visible hybridization in the Mo-$4d$ manifold with possible dual origin simillar to \ce{GaV4S8}. However, cluster motifs are not sufficient to lead to localized flatband formation. For example, in Figure~\ref{fig:dfts}c, \ce{Fe16Rb28Te32} exhibits \ce{Fe4} tetramers with an average cluster distance of 2.56~\AA, which is well below the Fe-Fe distance in pure metallic iron. However, several weakly dispersive Fe-3d bands appear around the Fermi level rather than a single isolated manifold. Though within a spin-polarized calculation a band gap opens, the band dispersion still remains (Figure \ref{fig:Fe16Rb28Te32_sp}). This illustrates a breakdown of the criterion which assumes that a metal--metal distance shorter than that of the pure metal will always produce isolated molecular orbitals (MOs) \cite{komleva_three-site_2020}. Thus, narrow bands in these materials can arise in different limits, ranging from relatively isolated cluster molecular orbitals to situations where cluster localization coexists with frustrated-lattice topology and stronger hybridization depending on cluster type, chemical environment, electron count and correlation strength \cite{kumari_molecular_2025}.

\begin{table}[tbp]
\centering
\caption{Top stable cluster materials with possible flatbands according to crytal net tight-binding model with relevant cluster and flat band properties. FB sub-lattice indicates transition metal sub-lattice responsible for the flat bands; No. FBs shows number of possible flat bands and FB lattice dimensionality indicates the dimensionality of the cluster sub-lattice.}
\resizebox{\textwidth}{!}{
\small
\begin{tabular}{l c c c c r c c c}
  \toprule
  Formula & Cluster sizes & Cluster space group & \makecell[c]{Cluster point\\groups} & \makecell[c]{Cluster\\dimensionality} & \makecell[c]{Min. avg.\\distance} & \makecell[c]{FB\\sublattice} & No. FBs & \makecell[c]{FB lattice\\dimensionality} \\
  \midrule
  Co$_{5}$SbO$_{8}$ & 3 & R$\bar{3}$m & $D_{3h}$ & 2 & 2.94 & Co & 1 & 2 \\
  Co$_{3}$(SnS)$_{2}$ & 3 & R$\bar{3}$m & $D_{3h}$ & 2 & 2.68 & Co & 1 & 2 \\
  Cu$_{3}$H$_{2}$(OF$_{2}$)$_{2}$ & 2 & P2/m & $D_{\infty h}$ & 3 & 2.98 & Cu & 2 & 1, 1 \\
  Ce$_{2}$B$_{2}$Ir$_{5}$ & 7, 3 & Cm & $D_{3d}$, $D_{3h}$ & 2 & 2.73 & Ir & 3 & 3 \\
  Li$_{7}$(Mo$_{3}$Se$_{4}$)$_{4}$ & 6, 6 & Cm & $D_{2h}$, $O_{h}$ & 3 & 2.66 & Mo & 4 & 1, 2 \\
  Ba$_{2}$Nb$_{15}$O$_{32}$ & 6 & R$\bar{3}$m & $D_{3d}$ & 2 & 2.84 & Nb & 1 & 2 \\
  Sn$_{2}$Rh$_{3}$S$_{2}$ & 3 & R$\bar{3}$m & $D_{3h}$ & 2 & 2.85 & Rh & 1 & 2 \\
  Sc$_{7}$NiBr$_{12}$ & 4 & R$\bar{3}$m & $C_{3v}$ & 3 & 2.44 & Sc & 4 & 1 \\
  Ta$_{3}$Al$_{2}$CoC & 2, 2, 2 & R3m & $C_{\infty v}$ & 2 & 2.98 & Co & 2 & 3 \\
  Sm$_{3}$Ti$_{3}$(SeO$_{4}$)$_{2}$ & 2, 2 & P2$_{1}$/m & $D_{\infty h}$ & 2 & 2.84 & Ti & 2 & 1 \\
  Li$_{2}$V$_{3}$SnO$_{8}$ & 3, 3 & Cmcm & $C_{2v}$ & 2 & 2.92 & V & 2 & 2, 2 \\
  V$_{4}$GaS$_{8}$ & 4 & Fm$\bar{3}$m & $T_{d}$ & 3 & 2.92 & V & 2 & 3 \\
  Zr$_{9}$S$_{2}$ & 2, 2 & I4/mmm & $D_{\infty h}$ & 3 & 2.91 & Zr & 4 & 3 \\
  \bottomrule
\end{tabular}
}
\label{fbt}
\end{table}

\section{Mixed-Metal Clusters}

\begin{figure}
\centering
\includegraphics[width=1\linewidth]{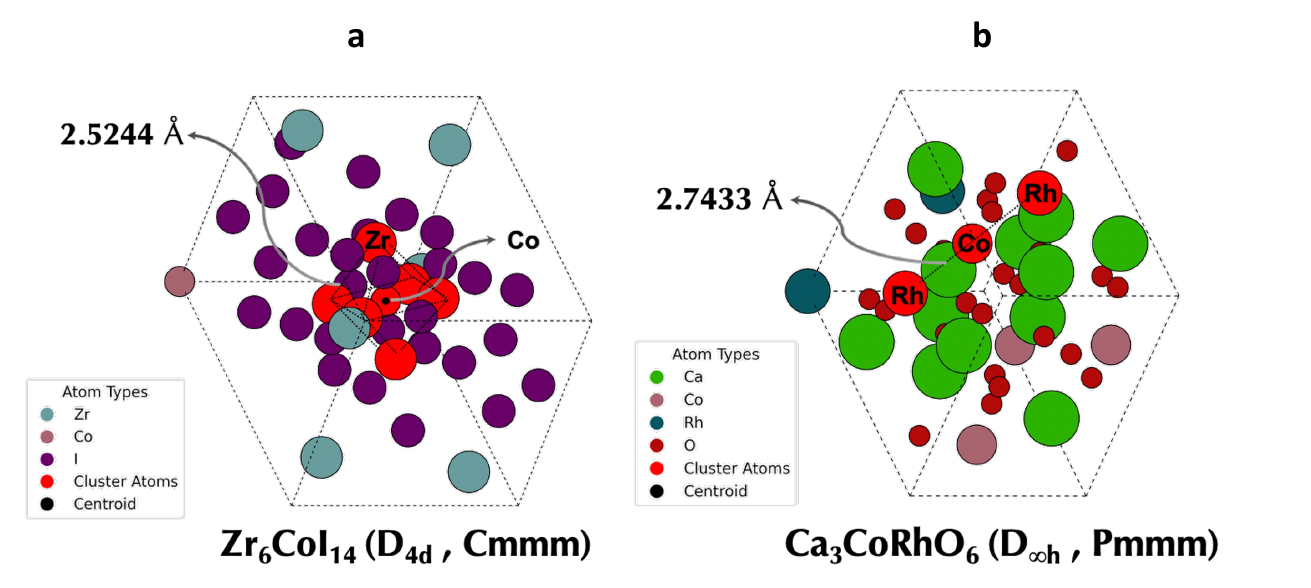}
\caption{Examples of mixed clusters, (a) \ce{Zr6CoI14} showing hexamer cluster of Zr with a hetero-atom (Co) inside the polyhedra and (b) \ce{Ca3CoRhO6} showing linear trimer with hetero-atom in the middle (Co instead of Rh).}
\label{fig:mixed}
\end{figure}

The cluster finder identified a large subset of materials with mixed-metal clusters - a particularly understudied direction.  We can also think of these mixed elements clusters in terms of doping e.g. \ce{Zr6CoI14} can be thought as if a Zr octahedra has been doped by Co at the cube center (Figure \ref{fig:mixed}a). This doping strategy has been used in \ce{Ba4RuMn2O10}, where by doping the trimers the material has led to a polar material, resulting in second harmonic generation \cite{skaggs_ba4rumn2o10_2024}.

We also found materials with extended metal-metal bonding, however with mixed metal types, e.g. \ce{Ca3CoRhO6} (Figure \ref{fig:mixed}b) which may hold quasi-one-dimensional Ising spin chains built from strongly coupled Rh-Co-Rh trimers of face-sharing octahedral RhO$_6$ and trigonal-prismatic CoO$_6$ polyhedra. Moreover, previous first-principles calculations and neutron diffraction establish that this chain geometry promotes substantial metal-metal overlap and a robust, highly anisotropic ferromagnetic coupling within the chains, while antiferromagnetic coupling between the chains arranged on a triangular lattice drives a partially disordered antiferromagnetic state below $T_1 \approx 90$~K with an additional low-temperature freezing near $T_2 \approx 30$~K \cite{paulose_spin-chain__2008, eyert_magnetic_2008, hardy_specific_2003}. While this is not a correlated electron molecular orbital material, it does highlight the relevance of the other types of correlated electron materials we've identified via our method.

\section{Electrochemical Control of Cluster Materials}

To identify materials within our dataset that are susceptible to electrochemical doping, we searched for common conducting ions in cathode materials, e.g., Li, Na, K, Mg, Zn, Ca, and Al, and also added relevant electrochemical information from the MP Battery Explorer for the resulting 470 available compounds. The varied geometries of transition metal cluster materials, and their different electron-crystal structure interplay may lead to more effective electrochemical control. Electrochemical doping may allow for the control of the correlated electron properties of this class of materials, as for example, in Mo$_3$-based kagome networks, including $Li_xScMo_3O_8$, where Li doping controls the spin state of the material\cite{wyckoff_electrochemical_2023}. This dynamic control is chemically attractive because redox-active metal-metal-bonded clusters can sometimes accommodate changes in electron count by distributing charge over cluster orbitals rather than localizing it on a single metal site, as demonstrated in lithiated \ce{LiScMo3O8} and in Chevrel-phase \ce{Mo6S8} intercalation chemistry \cite{richard_charge_2017}. Our work and methodology led to the identification of a wide variety of materials which can be electrochemically doped and controlled. 

\section{High-throughput Summary}

To identify if a material contains clusters, the "Cluster Finder" code first retrieves the crystal structure from Materials Project, builds a connectivity matrix for the TM sites and builds a graph from it. Graph representations not only capture this connectivity and provide appropiate features to model this kind of system, but also provide an opportunity to connect with other material properties (e.g. sublattice based flat bands, ion migration paths in battery materials) which can be represented in a simillar way. From the graph, using distance criteria (for example, the distances between the TM sites have to be $< 3\,\text{\AA}$) and splitting algorithms, we identify unique clusters. To classify the clusters, we determine their point group. Then, to understand how the clusters relate to each other through translations, we determine a centroid position for each unique cluster. Using the centroid positions as lattice points, we build a sub-lattice and determine its space group. Moreover, we determine the dimensionality of the spatial distribution of the centroids using Principal Component Analysis (Figure \ref{fig:flowchart}). More detail on this process will be discussed in the method section.

A high-throughput search was performed using our custom "Cluster Finder" code on a subset of Materials Project, on materials containing not more than four unique elements, with a total magnetization in the range of 0.01-10 $\mu B$. The total number of analyzed compounds were 34,548. From our experiments, these criteria, along with the parameters used to identify clusters, capture most of the cluster materials represented in the literature. However, if a broader subset of more varied and complex materials is to be analyzed with looser parameters, our code provides the necessary tools, including a built-in high-throughput search that can be run with parallel processing from the command line. The cluster-identification parameters can also be easily changed through a configuration file. More details can be found in the code itself.

\begin{figure}
\centering
\includegraphics[width=1\linewidth]{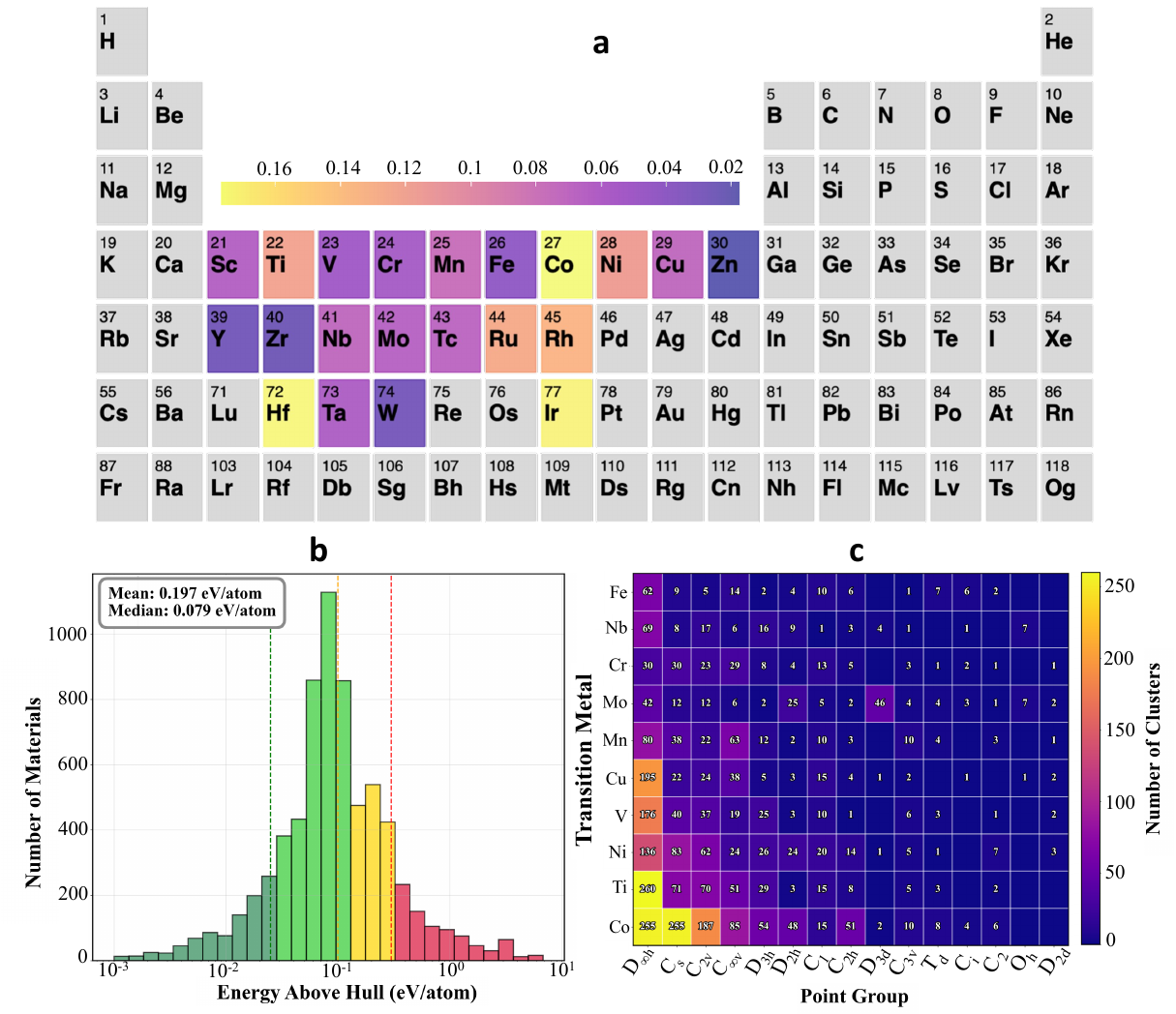}
\caption{Summary of the results of our high-throughput search using the Cluster Finder, (a) normalized likelihood of forming clusters by different transition metals in stable materials contained within Materials Project, (b) stability distribution of all the identified cluster materials in terms of energy above the hull with dark green indicating stable, light green metastable, yellow slightly unstable and red unstable cluster materials and (c) correlation heatmap showing the normalized likelihood of forming stable clusters with specific point groups by a transition metal.}
\label{fig:summary}
\end{figure}

Among the search space, the total number of compounds with possible clusters (after removing missing data and duplicates) was 5,306. For aditional analysis we further downselect materials based on the stability criterion that the material is within 0.1eV/atom from the convex Hull energy to account for errors in DFT and the possibility of synthesizing metastable materials (Figure \ref{fig:summary}b). The total count of stable/metastable compounds with only isolated cluster were 2,627. We also found a total of 984 mixed element clusters. Additionally, the total number of battery materials containing clusters was 1,590 (470 compounds with battery data from materials project).
Among the stable isolated clusters, the Co column of TMs in the periodic table were found to form more clusters, with Co forming the most. A deviation from this trend can be seen for the element Hf (Figure \ref{fig:summary}a). However, this may be less of a reflection of the tendency of the elements themselves, and more of a reflection of the compounds on materials project. Most transition metals show a strong preference for $D_{\infty h}$ symmetry (linear with inversion center), with decreasing numbers in lower symmetries including $C_s$ (mirror plane only), $C_{2v}$ (two-fold rotation) and $C_{\infty v}$ (linear without inversion, common for mixed element linear clusters) (Figure \ref{fig:summary}c).

Although the oxidation state and local coordination environment vary across the high-throughput dataset, there is often a useful naive correspondence between the preferred cluster type and the number of $d$ electrons per transition-metal site, $n$. In the simplest limit, a transition-metal dimer is favored for $d^1$ configurations, because the two metal sites can form a bonding molecular orbital that is fully occupied, analogous to the H$_2$ molecule. So for example $d^1$ configurations, with $d_{z^2}$ orbitals half filled, dimer formation becomes natural. The picture depends on the exact symmetry of the cluster and orbital filling, with for example triangular trimers preferring 6-8 electrons per trimer, with intermediate correlation strength best suited for cluster formation \cite{kumari_molecular_2025}. 
This electron-counting argument provides a qualitative explanation for why particular
transition metals show preferences for specific cluster sizes and point groups in our
high-throughput results. For example, Ti and V, which commonly occur in low-$d$-electron configurations, are especially frequent in simple cluster geometries: Ti has the largest number of $D_{\infty h}$ clusters in the dataset, with 260 occurrences, followed by substantial counts of $C_s$ and $C_{2v}$ motifs with 71 and 70 occurrences, respectively; V similarly shows a strong preference for $D_{\infty h}$ motifs, with 176 occurrences, together with notable $C_s$, $C_{2v}$, and $D_{3h}$ counts. Co and Ni show a broader distribution of point groups, consistent with their greater chemical flexibility and multiple accessible oxidation states: Co forms many $D_{\infty h}$, $C_s$, and $C_{2v}$ clusters, with 258, 255, and 187 occurrences, respectively, while Ni shows large counts in $D_{\infty h}$, $C_s$, and $C_{2v}$ motifs with 136, 83, and 62 occurrences. In contrast, Mo stands out for its relatively large number of $D_{3d}$ clusters, with 46 occurrences, comparable to or exceeding several of its simpler linear or low-symmetry motifs, consistent with the known tendency of Mo to form larger Mo$_{3n}$-type cluster units in Chevrel-like and related structures. These phases are built around $Mo_{3n}$, ranging from $Mo_{3}$ to much larger units like $Mo_{12}$ or $Mo_{18}$ in extended or shared frameworks. These clusters exhibit $D_{3d}$ point group symmetry due to a rhombohedral distortion of the ideal octahedron (which would be $O_{h}$). Moreover, highly symmetric point groups such as $O_{h}$ and $T_{d}$ can be seen for some specific TMs: Nb, Mo and Co, Fe respectively (Figure \ref{fig:summary}c). These examples show that the dominant point-group trends are not random: linear $D_{\infty h}$ motifs are common across many elements, triangular or distorted trimer-like motifs appear for early transition metals, and heavier cluster-forming elements such as Mo can stabilize larger, more symmetric cluster geometries. However, this correspondence should be viewed as a guiding principle rather than a strict rule, because real cluster formation also depends on ligand field splitting, metal--metal distance, covalency, oxidation state, lattice constraints, electron correlation
strength, and competing extended bonding motifs. 

\section{Methods}

\subsection{Cluster Identification}
\begin{figure}[tbp]
\centering
\begin{tikzpicture}[
  node distance=1cm,
  box/.style={rectangle, draw, text width=4.5cm, minimum height=1cm, align=center, rounded corners, fill=blue!10, font=\small},
  connbox/.style={rectangle, draw, text width=6cm, minimum height=1cm, align=center, rounded corners, fill=orange!10, font=\small},
  graphbox/.style={rectangle, draw, text width=4.5cm, minimum height=1cm, align=center, rounded corners, fill=green!10, font=\small},
  clusterbox/.style={rectangle, draw, text width=5.5cm, minimum height=1cm, align=center, rounded corners, fill=purple!10, font=\small},
  arrow/.style={-Stealth, thick},
  title/.style={font=\bfseries},
  ]

\node[title] at (0,0) (title) {Transition Metal Cluster Analysis Algorithm};

\node[box, below=0.8cm of title] (step1) {Step 1: Create Connectivity Matrix};
\node[graphbox, below=0.8cm of step1] (step2) {Step 2: Graph Representation};
\node[clusterbox, below=0.7cm of step2] (step3) {Step 3: Cluster Identification};
\node[box, below=1.2cm of step3] (step4) {Step 4: Analyze and Split Clusters};

\draw[arrow] (step1) -- (step2);
\draw[arrow] (step2) -- (step3);
\draw[arrow] (step3) -- (step4);

\node[connbox, right=0.6cm of step1] (conn_details) {
    \textbf{Connectivity Matrix $C$:} 
    $C_{ij} = \begin{cases} 
        1 & \text{if } d(s_i, s_j) < \text{cutoff} \\
        0 & \text{otherwise}
    \end{cases}$
};

\node[graphbox, right=0.6cm of step2] (graph_details) {
    \textbf{Graph $G = (V, E)$:} 
    $V = \{0,1,...,n-1\}$,
    $E = \{(i,j) | C_{ij} = 1\}$
};

\node[clusterbox, right=0.6cm of step3] (cluster_details) {
    \textbf{Clusters:} Connected components in $G$ where 
    size $\geq$ CLUSTER SIZE
};

\node[connbox, right=0.6cm of step4] (analyze_details) {
    \textbf{Avg Distance:} $\bar{d} = \frac{1}{|E|}\sum_{(i,j) \in E} d(s_i, s_j)$\\
    \textbf{Split if:} size $>$ CLUSTER SIZE+1 and
    $\bar{d}_{\text{parent}} - \bar{d}_{\text{sub}} > \text{threshold}$
};

\draw[arrow, dashed] (step1) -- (conn_details);
\draw[arrow, dashed] (step2) -- (graph_details);
\draw[arrow, dashed] (step3) -- (cluster_details);
\draw[arrow, dashed] (step4) -- (analyze_details);

\node[box, below=1.5cm of step4, xshift=-3cm] (split_algo) {
    \textbf{Split Algorithm}\\
};

\node[box, below=1.3cm of step4, xshift=3cm] (centroid_calc) {
    \textbf{Centroid Calculation:}\\
    1. $\vec{f}_i = \text{lattice}^{-1} \cdot \vec{r}_i$\\
    2. $\vec{f}_{\text{avg}} = \frac{1}{n}\sum_{i=1}^{n} \vec{f}_i$\\
    3. $\vec{c} = \text{lattice} \cdot \vec{f}_{\text{avg}}$
};

\draw[arrow] (step4) -- ++(0,-1.5) -| (split_algo);
\draw[arrow] (step4) -- ++(0,-1.5) -| (centroid_calc);

\begin{scope}[on background layer]
\node[fit=(title)(step1)(step2)(step3)(step4)(conn_details)(graph_details)
         (cluster_details)(analyze_details)(split_algo)(centroid_calc), 
      rectangle, draw, rounded corners, fill=gray!5, inner sep=0.3cm] {};
\end{scope}

\end{tikzpicture}
\caption{Mathematical flow and logic of transition metal cluster identification and analysis algorithm}
\label{fig:algorithm-overview}
\end{figure}

\begin{figure}
\centering
\includegraphics[width=0.85\linewidth]{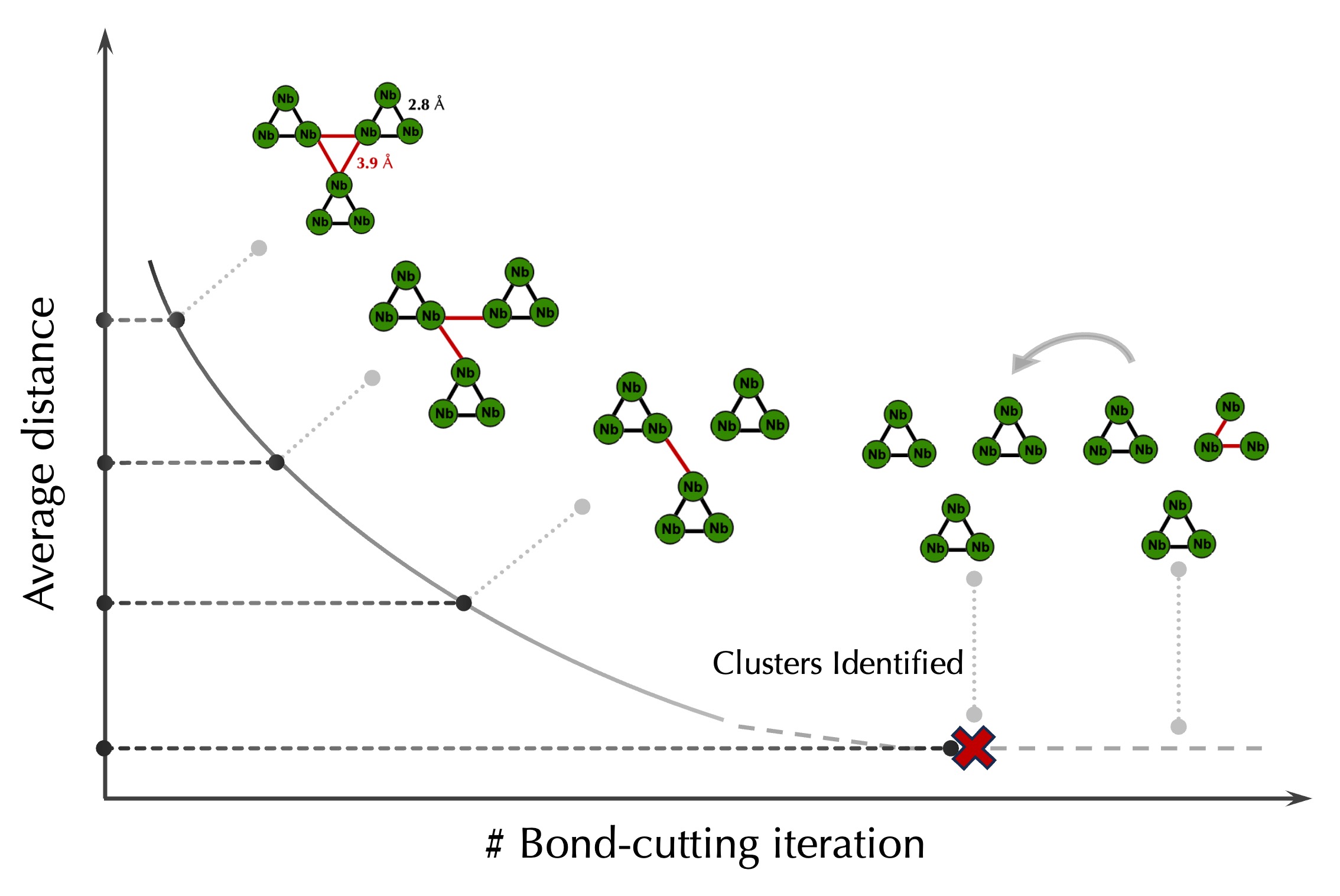}
\caption{Visualization of the cluster splitting process, as exemplified on Nb ions in a distorted kagome lattice. The algorithm starts with a cluster containing longer bonds (red lines), progressively cuts them, identifying them as  inter-cluster bonds, until the average bond lengths for the graphs stabilizes, identifying clusters with short average Nb-Nb bond lengths.}
\label{fig:split-cluster}
\end{figure}

To identify potential CEMO materials we have developed a custom algorithm as summarized in the diagram (Figure \ref{fig:algorithm-overview}). First, we begin by selecting compounds based on a predefined chemical composition criterion - for example, in the case of our high throughput search, compounds containing transition metals (TMs) such as Sc, Ti, V, Cr, Mn, Fe, Co, Ni, Cu, Zn, Y, Zr, Nb, Mo, Tc, Ru, Rh, Hf, Ta, W, Ir and anions such as B, C, N, O, P, S, Cl, Se, Br, Te, I with at most four distinct types of atoms with a total magnetization in the range of 0.01-10 $\mu B$. We then apply intial distance cutoff of $3\,\text{\AA}$ between TM sites to construct the connectivity matrix. Utilizing this matrix we build the initial parent graph where the TM sites form the nodes and distances between them form the edges. Using this initial graph we get the initial cluster guess and calculate the inital average distance at this point. Then, utilizing a moving cutoff scheme we cut the longest egde and calculate the average distances of the new child graphs (sub-clusters). If the new average metal-metal distance is lower than the initial (parent) average distance then we keep the split otherwise we keep the initial cluster (Figure \ref{fig:split-cluster}). The splitting process stops when the average bond length stabilizes, indicating the resulting graph accurately represents a physically meaningful cluster without unnecessarily fragmenting chemically coherent groups. This process is more reliable than simply relying on the assumption that a metal-metal distance shorter than that of the pure metal will always produce isolated molecular orbitals (MOs) \cite{komleva_three-site_2020}, and it is robust across a varied set of materials families. Finally, we calculate the center of point mass (centroid) of the clusters and use it to make sure the splitted clusters are well separated. The smallest cluster size is two and biggest is eight. Other properties and parameters of the cluster indentification process can be found in the configuration file used in our code. Following the initial identification of promising candidates based on these signatures, clusters within the identified materials are further characterized and their symmetry and lattice properties are analyzed in detail.

\subsection{Symmetry and Dimensionality Analysis}
Once TM clusters are identified, their internal symmetry is determined by calculating their point group symmetry centered on the centroid of the cluster pseudo-molecule. This local symmetry feature can capture potential couplings between electronic, magnetic, and structural degrees of freedom. After that, we utilize the centroid points to construct a sublattice and determine the space group of that sublattice which gives the translational relationship between the clusters in the 3D space. To build a proper sublattice, we label each centroid points in terms of its uniqueness. We utilize cluster properties e.g. element type, cluster size, point group, average distance and orientation (calculated using moment of inertia tensor) to determine the uniqueness. If any of these properties for two clusters are different we give them different/unique labels otherwise, we give them the same label. The extended sublattice structure formed by these clusters is further characterized by assigning each cluster unique spatial coordinates and using Principal Component Analysis (PCA) to determine the effective dimensionality-differentiating among three-dimensional frameworks, two-dimensional layers, one-dimensional chains, and zero-dimensional molecular like solids. Further discussion on the PCA method can be found in the supporting information.

\begin{figure}[tbp]
  \centering
  \begin{tikzpicture}[
      node distance=1.5cm,
      every node/.style={font=\footnotesize},
      box/.style={draw, rectangle, align=center, minimum width=2.8cm, minimum height=1.3cm},
      arrow/.style={-{Stealth}, thick},
    ]
    \node[box] (coords) {%
      \texttt{coords}\\
      $n\times3$\\[0.1em]
      $
      \begin{bmatrix} x_1 & y_1 & z_1 \\ \vdots & \vdots & \vdots \\ x_n & y_n & z_n
    \end{bmatrix}$};

    \node[box, right=2.2cm of coords] (centered) {%
      \texttt{centered}\\
      $n\times3$\\[0.1em]
      $
      \begin{bmatrix} x_1-\mu_x & y_1-\mu_y & z_1-\mu_z \\ \vdots & \vdots & \vdots \\ x_n-\mu_x & y_n-\mu_y & z_n-\mu_z
    \end{bmatrix}$};

    \node[box, right=2.8cm of centered] (svd) {%
      \texttt{SVD}\\[0.1em]
      $U: n\times3$\\
      $s: [s_0, s_1, s_2]$\\
    $V^T: 3\times3$};

    \node[box, below=1.8cm of svd] (s_norm) {%
      \texttt{s\_normalized}\\[0.1em]
    $[1,\, s_1/s_0,\, s_2/s_0]$};

    \node[box, left=2cm of s_norm] (classify) {%
      \texttt{Classify}\\[0.1em]
      ``0D'', ``1D'',\\
    ``2D'' or ``3D''};

    \draw[arrow] (coords) -- node[above=0.1cm, font=\scriptsize] {Subtract $\mu$} (centered);
    \draw[arrow] (centered) -- node[above=0.1cm, font=\scriptsize] {SVD: $U\,\mathrm{diag}(s)\,V^T$} (svd);
    \draw[arrow] (svd) -- node[right, font=\scriptsize] {Normalize $s$} (s_norm);
    \draw[arrow] (s_norm) -- node[above, font=\scriptsize] {Thresholds} (classify);
  \end{tikzpicture}
  \caption{SVD-based dimensionality classification process}
  \label{fig:svd-classification}
\end{figure}

For our dimentional analysis (Figure \ref{fig:svd-classification}), we extract 3D coordinates $(x, y, z)$ of $n$ centroid points from a large supercell constructed from our conventional cluster sublattice unit cell.
So, we have $n \times 3$, where $n$ is the number of sites (centroids), and 3 corresponds to the spatial dimensions which can be representated as,
\begin{equation}
\text{coords} =
\begin{bmatrix} x_1 & y_1 & z_1 \\ x_2 & y_2 & z_2 \\ \vdots & \vdots & \vdots \\ x_n & y_n & z_n
\end{bmatrix}
\end{equation}

Now, we compute the mean of each column $(\mu_x, \mu_y, \mu_z)$ and subtracts it from each coordinate to center the data around the origin, a prerequisite for PCA.

\[
\mu = \left[ \frac{\sum x_i}{n}, \frac{\sum y_i}{n}, \frac{\sum z_i}{n} \right], \quad \text{size } 1 \times 3.
\]
Through broadcasting we get,
\[
\text{centered} = \text{coords} - \frac{1}{n} \mathbf{1}_n \mu,
\]
where $\mathbf{1}_n$ is an $n \times 1$ vector of ones.
Now, we perform SVD as,
\[
\text{centered} = U \Sigma V^T
\]
Here, $V$ ($3 \times 3$) (columns of $V^T$ transposed) gives the principal directions in 3D space and $\Sigma$ (size 3 vector) is eigenvalues for those directions.
Then, we normalize the Singular Values as,
\[
\begin{aligned}
\text{s\_normalized}
  &= \left[ \frac{\sigma_0}{\sigma_{\max}},
             \frac{\sigma_1}{\sigma_{\max}},
             \frac{\sigma_2}{\sigma_{\max}} \right], \\
\sigma_{\max} &= \sigma_0.
\end{aligned}
\]
Thus,
\[
\text{s\_normalized} = \left[1, \frac{\sigma_1}{\sigma_0}, \frac{\sigma_2}{\sigma_0} \right], \quad \text{for } n \geq 3.
\]
This normalization facilitates comparison using thresholds, focusing on relative significance. The dimensionality is determined using the following criteria:
\begin{table}[htbp]
\centering
\begin{tabular}{l l}
\hline
\textbf{Dimensionality} & \textbf{Condition} \\
\hline
0D & $s[1] < 0.2$ and $s[2] < 0.2$ \\
1D & $1 - s[1] \leq 0.02$ and $s[2] < 0.5$ \\
2D & $\frac{s[2]}{s[1]} < 0.8$ \\
3D & $\frac{s[2]}{s[1]} \geq 0.8$ \\
\hline
\end{tabular}
\caption{Conditions for Dimensionality Classification Based on Normalized Singular Values}
\label{tab:dimensionality_conditions}
\end{table}

\section{Discussion}
Our work serves as a framework to classify and identify materials displaying metal clusters and extended metal-metal bonding in solids, identifying potential correlated electron molecular orbital (CEMO) materials, identifying their symmetry-related properties, as well as connected them to tools to analyze their potential flatband and battery properties. We have provided our database in an easy to use, interactive website. Our work can serve as a building block for the systematic study and discovery of this wide class of materials.

The additional electronic-structure analysis highlights that cluster motifs should not be treated only as local structural units, but also in relation to the extended lattice in which they are embedded. In some cases, such as \ce{Lu(NbCl3)6}, the narrow bands near the Fermi level are relatively well described by an isolated cluster molecular-orbital picture. In other cases, such as \ce{GaV4S8} and \ce{Li7(Mo3Se4)4}, cluster-localized states coexist with frustrated-lattice topology and stronger intercluster hybridization. This indicates that narrow bands in cluster materials can have a mixed origin, where molecular-orbital localization and lattice-derived flat-band mechanisms both contribute to the low-energy electronic structure.

At the same time, the examples discussed here show that short metal--metal distances alone are not sufficient to establish a CEMO description. The case of \ce{Fe16Rb28Te32} illustrates that even well-defined clusters with metal--metal distances shorter than those in the pure metal may still produce several weakly dispersive bands rather than a single isolated molecular-orbital manifold. Therefore, structural criteria are most useful as a first screening step, while the final classification of a candidate material requires consideration of electron count, cluster symmetry, chemical environment, hybridization, and correlation strength.

Finally, the identification of mixed-metal clusters and electrochemically active cluster materials expands the possible control parameters for this class of compounds. Mixed-metal clusters provide a route to tune local symmetry, electron filling, and magnetic exchange, while electrochemical doping offers a dynamic way to modify cluster-based electronic states. Together, these directions suggest that the cluster dataset developed here can be used not only to identify new CEMO candidates, but also to guide the design of materials in which cluster molecular orbitals, flat bands, magnetism, and ion insertion can be controlled in a systematic way.

\section{Code and Data availability}
The code to indentify, classify and analyze clusters is available at \href{https://github.com/RajbanulAkhond/Cluster_Finder}{Cluster Finder GitHub repo} with full CLI support. The cluster dataset containing flat band and electrochemical (battery) properties can be found here: \href{https://doi.org/10.24435/materialscloud:76-n2}{10.24435/materialscloud:76-n2}. To explore and download full dataset visit: \href{https://cluster-explorer.me/}{Cluster Explorer}.

\section{Acknowledgments}
This work was supported by the U.S. Department of Energy, Office of Science, Basic Energy Sciences, under the Early Career Research Program Award DE-SC0026069.This research was further supported in part by Lilly Endowment, Inc., through its support for the Indiana University Pervasive Technology Institute. The calculations presented in this work were performed on the Big Red 200 and Quartz supercomputer clusters at Indiana University. We acknowledge discussions with Tyrel McQueen, Shimin Zhang and Varsha Kumari.

\bibliographystyle{apsrev4-2}
\bibliography{references_v1}

\begin{thebibliography}{36}%
\makeatletter
\providecommand \@ifxundefined [1]{%
 \@ifx{#1\undefined}
}%
\providecommand \@ifnum [1]{%
 \ifnum #1\expandafter \@firstoftwo
 \else \expandafter \@secondoftwo
 \fi
}%
\providecommand \@ifx [1]{%
 \ifx #1\expandafter \@firstoftwo
 \else \expandafter \@secondoftwo
 \fi
}%
\providecommand \natexlab [1]{#1}%
\providecommand \enquote  [1]{``#1''}%
\providecommand \bibnamefont  [1]{#1}%
\providecommand \bibfnamefont [1]{#1}%
\providecommand \citenamefont [1]{#1}%
\providecommand \href@noop [0]{\@secondoftwo}%
\providecommand \href [0]{\begingroup \@sanitize@url \@href}%
\providecommand \@href[1]{\@@startlink{#1}\@@href}%
\providecommand \@@href[1]{\endgroup#1\@@endlink}%
\providecommand \@sanitize@url [0]{\catcode `\\12\catcode `\$12\catcode
  `\&12\catcode `\#12\catcode `\^12\catcode `\_12\catcode `\%12\relax}%
\providecommand \@@startlink[1]{}%
\providecommand \@@endlink[0]{}%
\providecommand \url  [0]{\begingroup\@sanitize@url \@url }%
\providecommand \@url [1]{\endgroup\@href {#1}{\urlprefix }}%
\providecommand \urlprefix  [0]{URL }%
\providecommand \Eprint [0]{\href }%
\providecommand \doibase [0]{https://doi.org/}%
\providecommand \selectlanguage [0]{\@gobble}%
\providecommand \bibinfo  [0]{\@secondoftwo}%
\providecommand \bibfield  [0]{\@secondoftwo}%
\providecommand \translation [1]{[#1]}%
\providecommand \BibitemOpen [0]{}%
\providecommand \bibitemStop [0]{}%
\providecommand \bibitemNoStop [0]{.\EOS\space}%
\providecommand \EOS [0]{\spacefactor3000\relax}%
\providecommand \BibitemShut  [1]{\csname bibitem#1\endcsname}%
\let\auto@bib@innerbib\@empty
\bibitem [{\citenamefont {Attfield}(2015)}]{attfield_orbital_2015}%
  \BibitemOpen
  \bibfield  {author} {\bibinfo {author} {\bibfnamefont {J.~P.}\ \bibnamefont
  {Attfield}},\ }\href {https://doi.org/10.1063/1.4913736} {\bibfield
  {journal} {\bibinfo  {journal} {APL Materials}\ }\textbf {\bibinfo {volume}
  {3}},\ \bibinfo {pages} {041510} (\bibinfo {year} {2015})}\BibitemShut
  {NoStop}%
\bibitem [{\citenamefont {Grytsiuk}\ \emph {et~al.}(2024)\citenamefont
  {Grytsiuk}, \citenamefont {Katsnelson}, \citenamefont {Loon},\ and\
  \citenamefont {Rösner}}]{grytsiuk_nb3cl8_2024}%
  \BibitemOpen
  \bibfield  {author} {\bibinfo {author} {\bibfnamefont {S.}~\bibnamefont
  {Grytsiuk}}, \bibinfo {author} {\bibfnamefont {M.~I.}\ \bibnamefont
  {Katsnelson}}, \bibinfo {author} {\bibfnamefont {E.~G. C. P.~v.}\
  \bibnamefont {Loon}},\ and\ \bibinfo {author} {\bibfnamefont
  {M.}~\bibnamefont {Rösner}},\ }\href
  {https://doi.org/10.1038/s41535-024-00619-5} {\bibfield  {journal} {\bibinfo
  {journal} {npj Quantum Materials}\ }\textbf {\bibinfo {volume} {9}},\
  \bibinfo {pages} {8} (\bibinfo {year} {2024})},\ \bibinfo {note} {publisher:
  Nature Publishing Group}\BibitemShut {NoStop}%
\bibitem [{\citenamefont {Gall}\ \emph {et~al.}(2013)\citenamefont {Gall},
  \citenamefont {Rahal Al~Orabi}, \citenamefont {Guizouarn},\ and\
  \citenamefont {Gougeon}}]{gall_synthesis_2013}%
  \BibitemOpen
  \bibfield  {author} {\bibinfo {author} {\bibfnamefont {P.}~\bibnamefont
  {Gall}}, \bibinfo {author} {\bibfnamefont {R.~A.}\ \bibnamefont {Rahal
  Al~Orabi}}, \bibinfo {author} {\bibfnamefont {T.}~\bibnamefont {Guizouarn}},\
  and\ \bibinfo {author} {\bibfnamefont {P.}~\bibnamefont {Gougeon}},\ }\href
  {https://doi.org/10.1016/j.jssc.2013.10.006} {\bibfield  {journal} {\bibinfo
  {journal} {Journal of Solid State Chemistry}\ }\textbf {\bibinfo {volume}
  {208}},\ \bibinfo {pages} {99} (\bibinfo {year} {2013})}\BibitemShut
  {NoStop}%
\bibitem [{\citenamefont {Mourigal}\ \emph {et~al.}(2014)\citenamefont
  {Mourigal}, \citenamefont {Fuhrman}, \citenamefont {Sheckelton},
  \citenamefont {Wartelle}, \citenamefont {Rodriguez-Rivera}, \citenamefont
  {Abernathy}, \citenamefont {McQueen},\ and\ \citenamefont
  {Broholm}}]{mourigal_molecular_magnet_2014}%
  \BibitemOpen
  \bibfield  {author} {\bibinfo {author} {\bibfnamefont {M.}~\bibnamefont
  {Mourigal}}, \bibinfo {author} {\bibfnamefont {W.~T.}\ \bibnamefont
  {Fuhrman}}, \bibinfo {author} {\bibfnamefont {J.~P.}\ \bibnamefont
  {Sheckelton}}, \bibinfo {author} {\bibfnamefont {A.}~\bibnamefont
  {Wartelle}}, \bibinfo {author} {\bibfnamefont {J.~A.}\ \bibnamefont
  {Rodriguez-Rivera}}, \bibinfo {author} {\bibfnamefont {D.~L.}\ \bibnamefont
  {Abernathy}}, \bibinfo {author} {\bibfnamefont {T.~M.}\ \bibnamefont
  {McQueen}},\ and\ \bibinfo {author} {\bibfnamefont {C.~L.}\ \bibnamefont
  {Broholm}},\ }\href {https://doi.org/10.1103/PhysRevLett.112.027202}
  {\bibfield  {journal} {\bibinfo  {journal} {Physical Review Letters}\
  }\textbf {\bibinfo {volume} {112}},\ \bibinfo {pages} {027202} (\bibinfo
  {year} {2014})},\ \bibinfo {note} {publisher: American Physical
  Society}\BibitemShut {NoStop}%
\bibitem [{\citenamefont {Marini}\ \emph {et~al.}(2021)\citenamefont {Marini},
  \citenamefont {Sanna}, \citenamefont {Pellegrini}, \citenamefont {Bersier},
  \citenamefont {Tosatti},\ and\ \citenamefont
  {Profeta}}]{marini_superconducting__2021}%
  \BibitemOpen
  \bibfield  {author} {\bibinfo {author} {\bibfnamefont {G.}~\bibnamefont
  {Marini}}, \bibinfo {author} {\bibfnamefont {A.}~\bibnamefont {Sanna}},
  \bibinfo {author} {\bibfnamefont {C.}~\bibnamefont {Pellegrini}}, \bibinfo
  {author} {\bibfnamefont {C.}~\bibnamefont {Bersier}}, \bibinfo {author}
  {\bibfnamefont {E.}~\bibnamefont {Tosatti}},\ and\ \bibinfo {author}
  {\bibfnamefont {G.}~\bibnamefont {Profeta}},\ }\href
  {https://doi.org/10.1103/PhysRevB.103.144507} {\bibfield  {journal} {\bibinfo
   {journal} {Physical Review B}\ }\textbf {\bibinfo {volume} {103}},\ \bibinfo
  {pages} {144507} (\bibinfo {year} {2021})},\ \bibinfo {note} {publisher:
  American Physical Society}\BibitemShut {NoStop}%
\bibitem [{\citenamefont {Kelly}\ \emph {et~al.}(2019)\citenamefont {Kelly},
  \citenamefont {Tran},\ and\ \citenamefont
  {McQueen}}]{kelly_nonpolar--polar_2019}%
  \BibitemOpen
  \bibfield  {author} {\bibinfo {author} {\bibfnamefont {Z.~A.}\ \bibnamefont
  {Kelly}}, \bibinfo {author} {\bibfnamefont {T.~T.}\ \bibnamefont {Tran}},\
  and\ \bibinfo {author} {\bibfnamefont {T.~M.}\ \bibnamefont {McQueen}},\
  }\href {https://doi.org/10.1021/acs.inorgchem.9b01110} {\bibfield  {journal}
  {\bibinfo  {journal} {Inorganic Chemistry}\ }\textbf {\bibinfo {volume}
  {58}},\ \bibinfo {pages} {11941} (\bibinfo {year} {2019})},\ \bibinfo {note}
  {publisher: American Chemical Society}\BibitemShut {NoStop}%
\bibitem [{\citenamefont {Kumari}\ \emph {et~al.}(2025)\citenamefont {Kumari},
  \citenamefont {Bauer},\ and\ \citenamefont
  {Georgescu}}]{kumari_molecular_2025}%
  \BibitemOpen
  \bibfield  {author} {\bibinfo {author} {\bibfnamefont {V.}~\bibnamefont
  {Kumari}}, \bibinfo {author} {\bibfnamefont {J.}~\bibnamefont {Bauer}},\ and\
  \bibinfo {author} {\bibfnamefont {A.~B.}\ \bibnamefont {Georgescu}},\
  }\bibfield  {journal} {\bibinfo  {journal} {Journal of Materials Chemistry
  C}\ }\href {https://doi.org/10.1039/D5TC01981H} {10.1039/D5TC01981H}
  (\bibinfo {year} {2025}),\ \bibinfo {note} {publisher: The Royal Society of
  Chemistry}\BibitemShut {NoStop}%
\bibitem [{\citenamefont {Wu}\ \emph {et~al.}(2022)\citenamefont {Wu},
  \citenamefont {Wang}, \citenamefont {Xu}, \citenamefont {Sivakumar},
  \citenamefont {Pasco}, \citenamefont {Filippozzi}, \citenamefont {Parkin},
  \citenamefont {Zeng}, \citenamefont {McQueen},\ and\ \citenamefont
  {Ali}}]{wu_field-free_2022}%
  \BibitemOpen
  \bibfield  {author} {\bibinfo {author} {\bibfnamefont {H.}~\bibnamefont
  {Wu}}, \bibinfo {author} {\bibfnamefont {Y.}~\bibnamefont {Wang}}, \bibinfo
  {author} {\bibfnamefont {Y.}~\bibnamefont {Xu}}, \bibinfo {author}
  {\bibfnamefont {P.~K.}\ \bibnamefont {Sivakumar}}, \bibinfo {author}
  {\bibfnamefont {C.}~\bibnamefont {Pasco}}, \bibinfo {author} {\bibfnamefont
  {U.}~\bibnamefont {Filippozzi}}, \bibinfo {author} {\bibfnamefont {S.~S.~P.}\
  \bibnamefont {Parkin}}, \bibinfo {author} {\bibfnamefont {Y.-J.}\
  \bibnamefont {Zeng}}, \bibinfo {author} {\bibfnamefont {T.}~\bibnamefont
  {McQueen}},\ and\ \bibinfo {author} {\bibfnamefont {M.~N.}\ \bibnamefont
  {Ali}},\ }\href {https://doi.org/10.1038/s41586-022-04504-8} {\bibfield
  {journal} {\bibinfo  {journal} {Nature}\ }\textbf {\bibinfo {volume} {604}},\
  \bibinfo {pages} {653} (\bibinfo {year} {2022})},\ \bibinfo {note}
  {publisher: Nature Publishing Group}\BibitemShut {NoStop}%
\bibitem [{\citenamefont {Nikolaev}\ \emph {et~al.}(2021)\citenamefont
  {Nikolaev}, \citenamefont {Solovyev},\ and\ \citenamefont
  {Streltsov}}]{nikolaev_quantum_2021}%
  \BibitemOpen
  \bibfield  {author} {\bibinfo {author} {\bibfnamefont {S.~A.}\ \bibnamefont
  {Nikolaev}}, \bibinfo {author} {\bibfnamefont {I.~V.}\ \bibnamefont
  {Solovyev}},\ and\ \bibinfo {author} {\bibfnamefont {S.~V.}\ \bibnamefont
  {Streltsov}},\ }\href {https://doi.org/10.1038/s41535-021-00316-7} {\bibfield
   {journal} {\bibinfo  {journal} {npj Quantum Materials}\ }\textbf {\bibinfo
  {volume} {6}},\ \bibinfo {pages} {25} (\bibinfo {year} {2021})},\ \bibinfo
  {note} {publisher: Nature Publishing Group}\BibitemShut {NoStop}%
\bibitem [{\citenamefont {Haraguchi}\ \emph {et~al.}(2015)\citenamefont
  {Haraguchi}, \citenamefont {Michioka}, \citenamefont {Imai}, \citenamefont
  {Ueda},\ and\ \citenamefont {Yoshimura}}]{haraguchi_spin--liquid_2015}%
  \BibitemOpen
  \bibfield  {author} {\bibinfo {author} {\bibfnamefont {Y.}~\bibnamefont
  {Haraguchi}}, \bibinfo {author} {\bibfnamefont {C.}~\bibnamefont {Michioka}},
  \bibinfo {author} {\bibfnamefont {M.}~\bibnamefont {Imai}}, \bibinfo {author}
  {\bibfnamefont {H.}~\bibnamefont {Ueda}},\ and\ \bibinfo {author}
  {\bibfnamefont {K.}~\bibnamefont {Yoshimura}},\ }\href
  {https://doi.org/10.1103/PhysRevB.92.014409} {\bibfield  {journal} {\bibinfo
  {journal} {Physical Review B}\ }\textbf {\bibinfo {volume} {92}},\ \bibinfo
  {pages} {014409} (\bibinfo {year} {2015})},\ \bibinfo {note} {publisher:
  American Physical Society}\BibitemShut {NoStop}%
\bibitem [{\citenamefont {Wyckoff}\ \emph {et~al.}(2023)\citenamefont
  {Wyckoff}, \citenamefont {Kautzsch}, \citenamefont {Kaufman}, \citenamefont
  {Ortiz}, \citenamefont {Kallistova}, \citenamefont {Pokharel}, \citenamefont
  {Liu}, \citenamefont {Taddei}, \citenamefont {Wiaderek}, \citenamefont
  {Lapidus}, \citenamefont {Wilson}, \citenamefont {Van~der Ven},\ and\
  \citenamefont {Seshadri}}]{wyckoff_electrochemical_2023}%
  \BibitemOpen
  \bibfield  {author} {\bibinfo {author} {\bibfnamefont {K.~E.}\ \bibnamefont
  {Wyckoff}}, \bibinfo {author} {\bibfnamefont {L.}~\bibnamefont {Kautzsch}},
  \bibinfo {author} {\bibfnamefont {J.~L.}\ \bibnamefont {Kaufman}}, \bibinfo
  {author} {\bibfnamefont {B.~R.}\ \bibnamefont {Ortiz}}, \bibinfo {author}
  {\bibfnamefont {A.}~\bibnamefont {Kallistova}}, \bibinfo {author}
  {\bibfnamefont {G.}~\bibnamefont {Pokharel}}, \bibinfo {author}
  {\bibfnamefont {J.}~\bibnamefont {Liu}}, \bibinfo {author} {\bibfnamefont
  {K.~M.}\ \bibnamefont {Taddei}}, \bibinfo {author} {\bibfnamefont {K.~M.}\
  \bibnamefont {Wiaderek}}, \bibinfo {author} {\bibfnamefont {S.~H.}\
  \bibnamefont {Lapidus}}, \bibinfo {author} {\bibfnamefont {S.~D.}\
  \bibnamefont {Wilson}}, \bibinfo {author} {\bibfnamefont {A.}~\bibnamefont
  {Van~der Ven}},\ and\ \bibinfo {author} {\bibfnamefont {R.}~\bibnamefont
  {Seshadri}},\ }\href {https://doi.org/10.1021/acs.chemmater.3c00202}
  {\bibfield  {journal} {\bibinfo  {journal} {Chemistry of Materials}\ }\textbf
  {\bibinfo {volume} {35}},\ \bibinfo {pages} {4945} (\bibinfo {year}
  {2023})},\ \bibinfo {note} {publisher: American Chemical Society}\BibitemShut
  {NoStop}%
\bibitem [{\citenamefont {Mei}\ \emph {et~al.}(2017)\citenamefont {Mei},
  \citenamefont {Xu}, \citenamefont {Wei}, \citenamefont {Liu}, \citenamefont
  {Li}, \citenamefont {Ma},\ and\ \citenamefont {Dou}}]{mei_chevrel_2017}%
  \BibitemOpen
  \bibfield  {author} {\bibinfo {author} {\bibfnamefont {L.}~\bibnamefont
  {Mei}}, \bibinfo {author} {\bibfnamefont {J.}~\bibnamefont {Xu}}, \bibinfo
  {author} {\bibfnamefont {Z.}~\bibnamefont {Wei}}, \bibinfo {author}
  {\bibfnamefont {H.}~\bibnamefont {Liu}}, \bibinfo {author} {\bibfnamefont
  {Y.}~\bibnamefont {Li}}, \bibinfo {author} {\bibfnamefont {J.}~\bibnamefont
  {Ma}},\ and\ \bibinfo {author} {\bibfnamefont {S.}~\bibnamefont {Dou}},\
  }\href {https://doi.org/10.1002/smll.201701441} {\bibfield  {journal}
  {\bibinfo  {journal} {Small}\ }\textbf {\bibinfo {volume} {13}},\ \bibinfo
  {pages} {1701441} (\bibinfo {year} {2017})},\ \bibinfo {note} {publisher:
  John Wiley \& Sons, Ltd}\BibitemShut {NoStop}%
\bibitem [{\citenamefont {Banerjee}\ \emph {et~al.}(2024)\citenamefont
  {Banerjee}, \citenamefont {Ghosh}, \citenamefont {Kataria},\ and\
  \citenamefont {Sundaresan}}]{banerjee_evidence__2024}%
  \BibitemOpen
  \bibfield  {author} {\bibinfo {author} {\bibfnamefont {S.}~\bibnamefont
  {Banerjee}}, \bibinfo {author} {\bibfnamefont {S.}~\bibnamefont {Ghosh}},
  \bibinfo {author} {\bibfnamefont {A.}~\bibnamefont {Kataria}},\ and\ \bibinfo
  {author} {\bibfnamefont {A.}~\bibnamefont {Sundaresan}},\ }\href
  {https://doi.org/10.1103/PhysRevB.110.014512} {\bibfield  {journal} {\bibinfo
   {journal} {Physical Review B}\ }\textbf {\bibinfo {volume} {110}},\ \bibinfo
  {pages} {014512} (\bibinfo {year} {2024})},\ \bibinfo {note} {publisher:
  American Physical Society}\BibitemShut {NoStop}%
\bibitem [{\citenamefont {Gosławska}\ and\ \citenamefont
  {Matlak}(1998)}]{goslawska_reentrant__1998}%
  \BibitemOpen
  \bibfield  {author} {\bibinfo {author} {\bibfnamefont {E.}~\bibnamefont
  {Gosławska}}\ and\ \bibinfo {author} {\bibfnamefont {M.}~\bibnamefont
  {Matlak}},\ }\href
  {https://doi.org/10.1002/(SICI)1521-3951(199806)207:2<469::AID-PSSB469>3.0.CO;2-X}
  {\bibfield  {journal} {\bibinfo  {journal} {physica status solidi (b)}\
  }\textbf {\bibinfo {volume} {207}},\ \bibinfo {pages} {469} (\bibinfo {year}
  {1998})}\BibitemShut {NoStop}%
\bibitem [{\citenamefont {Xu}\ and\ \citenamefont
  {Xiang}(2015)}]{xu_unusual_2015}%
  \BibitemOpen
  \bibfield  {author} {\bibinfo {author} {\bibfnamefont {K.}~\bibnamefont
  {Xu}}\ and\ \bibinfo {author} {\bibfnamefont {H.~J.}\ \bibnamefont {Xiang}},\
  }\href {https://doi.org/10.1103/PhysRevB.92.121112} {\bibfield  {journal}
  {\bibinfo  {journal} {Physical Review B}\ }\textbf {\bibinfo {volume} {92}},\
  \bibinfo {pages} {121112} (\bibinfo {year} {2015})}\BibitemShut {NoStop}%
\bibitem [{\citenamefont {Singh}\ \emph {et~al.}(2014)\citenamefont {Singh},
  \citenamefont {Simon}, \citenamefont {Cannuccia}, \citenamefont {Lepetit},
  \citenamefont {Corraze}, \citenamefont {Janod},\ and\ \citenamefont
  {Cario}}]{singh_orbital-ordering-driven__2014}%
  \BibitemOpen
  \bibfield  {author} {\bibinfo {author} {\bibfnamefont {K.}~\bibnamefont
  {Singh}}, \bibinfo {author} {\bibfnamefont {C.}~\bibnamefont {Simon}},
  \bibinfo {author} {\bibfnamefont {E.}~\bibnamefont {Cannuccia}}, \bibinfo
  {author} {\bibfnamefont {M.-B.}\ \bibnamefont {Lepetit}}, \bibinfo {author}
  {\bibfnamefont {B.}~\bibnamefont {Corraze}}, \bibinfo {author} {\bibfnamefont
  {E.}~\bibnamefont {Janod}},\ and\ \bibinfo {author} {\bibfnamefont
  {L.}~\bibnamefont {Cario}},\ }\href
  {https://doi.org/10.1103/PhysRevLett.113.137602} {\bibfield  {journal}
  {\bibinfo  {journal} {Physical Review Letters}\ }\textbf {\bibinfo {volume}
  {113}},\ \bibinfo {pages} {137602} (\bibinfo {year} {2014})},\ \bibinfo
  {note} {publisher: American Physical Society}\BibitemShut {NoStop}%
\bibitem [{\citenamefont {Ruff}\ \emph {et~al.}(2015)\citenamefont {Ruff},
  \citenamefont {Widmann}, \citenamefont {Lunkenheimer}, \citenamefont
  {Tsurkan}, \citenamefont {Bordács}, \citenamefont {Kézsmárki},\ and\
  \citenamefont {Loidl}}]{ruff_multiferroicity_2015}%
  \BibitemOpen
  \bibfield  {author} {\bibinfo {author} {\bibfnamefont {E.}~\bibnamefont
  {Ruff}}, \bibinfo {author} {\bibfnamefont {S.}~\bibnamefont {Widmann}},
  \bibinfo {author} {\bibfnamefont {P.}~\bibnamefont {Lunkenheimer}}, \bibinfo
  {author} {\bibfnamefont {V.}~\bibnamefont {Tsurkan}}, \bibinfo {author}
  {\bibfnamefont {S.}~\bibnamefont {Bordács}}, \bibinfo {author}
  {\bibfnamefont {I.}~\bibnamefont {Kézsmárki}},\ and\ \bibinfo {author}
  {\bibfnamefont {A.}~\bibnamefont {Loidl}},\ }\href
  {https://doi.org/10.1126/sciadv.1500916} {\bibfield  {journal} {\bibinfo
  {journal} {Science Advances}\ }\textbf {\bibinfo {volume} {1}},\ \bibinfo
  {pages} {e1500916} (\bibinfo {year} {2015})},\ \bibinfo {note} {publisher:
  American Association for the Advancement of Science}\BibitemShut {NoStop}%
\bibitem [{\citenamefont {Dorolti}\ \emph {et~al.}(2010)\citenamefont
  {Dorolti}, \citenamefont {Cario}, \citenamefont {Corraze}, \citenamefont
  {Janod}, \citenamefont {Vaju}, \citenamefont {Koo}, \citenamefont {Kan},\
  and\ \citenamefont {Whangbo}}]{dorolti_half-metallic_2010}%
  \BibitemOpen
  \bibfield  {author} {\bibinfo {author} {\bibfnamefont {E.}~\bibnamefont
  {Dorolti}}, \bibinfo {author} {\bibfnamefont {L.}~\bibnamefont {Cario}},
  \bibinfo {author} {\bibfnamefont {B.}~\bibnamefont {Corraze}}, \bibinfo
  {author} {\bibfnamefont {E.}~\bibnamefont {Janod}}, \bibinfo {author}
  {\bibfnamefont {C.}~\bibnamefont {Vaju}}, \bibinfo {author} {\bibfnamefont
  {H.-J.}\ \bibnamefont {Koo}}, \bibinfo {author} {\bibfnamefont
  {E.}~\bibnamefont {Kan}},\ and\ \bibinfo {author} {\bibfnamefont {M.-H.}\
  \bibnamefont {Whangbo}},\ }\href {https://doi.org/10.1021/ja908128b}
  {\bibfield  {journal} {\bibinfo  {journal} {Journal of the American Chemical
  Society}\ }\textbf {\bibinfo {volume} {132}},\ \bibinfo {pages} {5704}
  (\bibinfo {year} {2010})},\ \bibinfo {note} {published March 31,
  2010}\BibitemShut {NoStop}%
\bibitem [{\citenamefont {Takegami}\ \emph {et~al.}(2025)\citenamefont
  {Takegami}, \citenamefont {Kitou}, \citenamefont {Tokunaga}, \citenamefont
  {Yoshimura}, \citenamefont {Tsuei}, \citenamefont {Arima},\ and\
  \citenamefont {Mizokawa}}]{opticalspinel}%
  \BibitemOpen
  \bibfield  {author} {\bibinfo {author} {\bibfnamefont {D.}~\bibnamefont
  {Takegami}}, \bibinfo {author} {\bibfnamefont {S.}~\bibnamefont {Kitou}},
  \bibinfo {author} {\bibfnamefont {Y.}~\bibnamefont {Tokunaga}}, \bibinfo
  {author} {\bibfnamefont {M.}~\bibnamefont {Yoshimura}}, \bibinfo {author}
  {\bibfnamefont {K.-D.}\ \bibnamefont {Tsuei}}, \bibinfo {author}
  {\bibfnamefont {T.-h.}\ \bibnamefont {Arima}},\ and\ \bibinfo {author}
  {\bibfnamefont {T.}~\bibnamefont {Mizokawa}},\ }\href
  {https://doi.org/10.1103/PhysRevB.111.195155} {\bibfield  {journal} {\bibinfo
   {journal} {Phys. Rev. B}\ }\textbf {\bibinfo {volume} {111}},\ \bibinfo
  {pages} {195155} (\bibinfo {year} {2025})}\BibitemShut {NoStop}%
\bibitem [{\citenamefont {Aretz}\ \emph {et~al.}(2025)\citenamefont {Aretz},
  \citenamefont {Grytsiuk}, \citenamefont {Liu}, \citenamefont {Feraco},
  \citenamefont {Knekna}, \citenamefont {Waseem}, \citenamefont {Dan},
  \citenamefont {Bianchi}, \citenamefont {Hofmann}, \citenamefont {Ali},
  \citenamefont {Katsnelson}, \citenamefont {Grubi\ifmmode \check{s}\else
  \v{s}\fi{}i\ifmmode \acute{c}\else \'{c}\fi{}-\ifmmode~\check{C}\else
  \v{C}\fi{}abo}, \citenamefont {Strand}, \citenamefont {van Loon},\ and\
  \citenamefont {R\"osner}}]{hugomalte}%
  \BibitemOpen
  \bibfield  {author} {\bibinfo {author} {\bibfnamefont {J.}~\bibnamefont
  {Aretz}}, \bibinfo {author} {\bibfnamefont {S.}~\bibnamefont {Grytsiuk}},
  \bibinfo {author} {\bibfnamefont {X.}~\bibnamefont {Liu}}, \bibinfo {author}
  {\bibfnamefont {G.}~\bibnamefont {Feraco}}, \bibinfo {author} {\bibfnamefont
  {C.}~\bibnamefont {Knekna}}, \bibinfo {author} {\bibfnamefont
  {M.}~\bibnamefont {Waseem}}, \bibinfo {author} {\bibfnamefont
  {Z.}~\bibnamefont {Dan}}, \bibinfo {author} {\bibfnamefont {M.}~\bibnamefont
  {Bianchi}}, \bibinfo {author} {\bibfnamefont {P.}~\bibnamefont {Hofmann}},
  \bibinfo {author} {\bibfnamefont {M.~N.}\ \bibnamefont {Ali}}, \bibinfo
  {author} {\bibfnamefont {M.~I.}\ \bibnamefont {Katsnelson}}, \bibinfo
  {author} {\bibfnamefont {A.}~\bibnamefont {Grubi\ifmmode \check{s}\else
  \v{s}\fi{}i\ifmmode \acute{c}\else \'{c}\fi{}-\ifmmode~\check{C}\else
  \v{C}\fi{}abo}}, \bibinfo {author} {\bibfnamefont {H.~U.~R.}\ \bibnamefont
  {Strand}}, \bibinfo {author} {\bibfnamefont {E.~G. C.~P.}\ \bibnamefont {van
  Loon}},\ and\ \bibinfo {author} {\bibfnamefont {M.}~\bibnamefont
  {R\"osner}},\ }\href {https://doi.org/10.1103/wr7w-nfhg} {\bibfield
  {journal} {\bibinfo  {journal} {Phys. Rev. X}\ }\textbf {\bibinfo {volume}
  {15}},\ \bibinfo {pages} {041042} (\bibinfo {year} {2025})}\BibitemShut
  {NoStop}%
\bibitem [{\citenamefont {Kim}\ \emph {et~al.}(2020)\citenamefont {Kim},
  \citenamefont {Haule},\ and\ \citenamefont
  {Vanderbilt}}]{kim_molecular_2020}%
  \BibitemOpen
  \bibfield  {author} {\bibinfo {author} {\bibfnamefont {H.-S.}\ \bibnamefont
  {Kim}}, \bibinfo {author} {\bibfnamefont {K.}~\bibnamefont {Haule}},\ and\
  \bibinfo {author} {\bibfnamefont {D.}~\bibnamefont {Vanderbilt}},\ }\href
  {https://doi.org/10.1103/PhysRevB.102.081105} {\bibfield  {journal} {\bibinfo
   {journal} {Physical Review B}\ }\textbf {\bibinfo {volume} {102}},\ \bibinfo
  {pages} {081105} (\bibinfo {year} {2020})},\ \bibinfo {note} {publisher:
  American Physical Society}\BibitemShut {NoStop}%
\bibitem [{\citenamefont {Bersuker}(2013)}]{bersuker2013pseudo}%
  \BibitemOpen
  \bibfield  {author} {\bibinfo {author} {\bibfnamefont {I.~B.}\ \bibnamefont
  {Bersuker}},\ }\href@noop {} {\bibfield  {journal} {\bibinfo  {journal}
  {Chemical Reviews}\ }\textbf {\bibinfo {volume} {113}},\ \bibinfo {pages}
  {1351} (\bibinfo {year} {2013})}\BibitemShut {NoStop}%
\bibitem [{\citenamefont {Park}\ \emph {et~al.}(2020)\citenamefont {Park},
  \citenamefont {Sim}, \citenamefont {Jeong}, \citenamefont {Mishra},
  \citenamefont {Han},\ and\ \citenamefont {Lee}}]{park_pressure-induced_2020}%
  \BibitemOpen
  \bibfield  {author} {\bibinfo {author} {\bibfnamefont {M.~J.}\ \bibnamefont
  {Park}}, \bibinfo {author} {\bibfnamefont {G.}~\bibnamefont {Sim}}, \bibinfo
  {author} {\bibfnamefont {M.~Y.}\ \bibnamefont {Jeong}}, \bibinfo {author}
  {\bibfnamefont {A.}~\bibnamefont {Mishra}}, \bibinfo {author} {\bibfnamefont
  {M.~J.}\ \bibnamefont {Han}},\ and\ \bibinfo {author} {\bibfnamefont
  {S.}~\bibnamefont {Lee}},\ }\href {https://doi.org/10.1038/s41535-020-0246-0}
  {\bibfield  {journal} {\bibinfo  {journal} {npj Quantum Materials}\ }\textbf
  {\bibinfo {volume} {5}},\ \bibinfo {pages} {41} (\bibinfo {year} {2020})},\
  \bibinfo {note} {publisher: Nature Publishing Group}\BibitemShut {NoStop}%
\bibitem [{\citenamefont {Khomskii}\ and\ \citenamefont
  {Streltsov}(2021)}]{khomskii_orbital_2021}%
  \BibitemOpen
  \bibfield  {author} {\bibinfo {author} {\bibfnamefont {D.~I.}\ \bibnamefont
  {Khomskii}}\ and\ \bibinfo {author} {\bibfnamefont {S.~V.}\ \bibnamefont
  {Streltsov}},\ }\href {https://doi.org/10.1021/acs.chemrev.0c00579}
  {\bibfield  {journal} {\bibinfo  {journal} {Chemical Reviews}\ }\textbf
  {\bibinfo {volume} {121}},\ \bibinfo {pages} {2992} (\bibinfo {year}
  {2021})},\ \bibinfo {note} {arXiv:2006.05920 [cond-mat]}\BibitemShut
  {NoStop}%
\bibitem [{\citenamefont {Zhong}\ \emph {et~al.}(2020)\citenamefont {Zhong},
  \citenamefont {Guo}, \citenamefont {Nguyen},\ and\ \citenamefont
  {Cava}}]{zhong_frustrated_2020}%
  \BibitemOpen
  \bibfield  {author} {\bibinfo {author} {\bibfnamefont {R.}~\bibnamefont
  {Zhong}}, \bibinfo {author} {\bibfnamefont {S.}~\bibnamefont {Guo}}, \bibinfo
  {author} {\bibfnamefont {L.~T.}\ \bibnamefont {Nguyen}},\ and\ \bibinfo
  {author} {\bibfnamefont {R.~J.}\ \bibnamefont {Cava}},\ }\href
  {https://doi.org/10.1103/PhysRevB.102.224430} {\bibfield  {journal} {\bibinfo
   {journal} {Physical Review B}\ }\textbf {\bibinfo {volume} {102}},\ \bibinfo
  {pages} {224430} (\bibinfo {year} {2020})},\ \bibinfo {note} {publisher:
  American Physical Society}\BibitemShut {NoStop}%
\bibitem [{\citenamefont {Cao}\ \emph {et~al.}(2020)\citenamefont {Cao},
  \citenamefont {Zheng}, \citenamefont {Zhao}, \citenamefont {Ni},
  \citenamefont {Pocs}, \citenamefont {Zhang}, \citenamefont {Ye},
  \citenamefont {Hoffmann}, \citenamefont {Wang}, \citenamefont {Lee},
  \citenamefont {Hermele},\ and\ \citenamefont {Kimchi}}]{cao_quantum_2020}%
  \BibitemOpen
  \bibfield  {author} {\bibinfo {author} {\bibfnamefont {G.}~\bibnamefont
  {Cao}}, \bibinfo {author} {\bibfnamefont {H.}~\bibnamefont {Zheng}}, \bibinfo
  {author} {\bibfnamefont {H.}~\bibnamefont {Zhao}}, \bibinfo {author}
  {\bibfnamefont {Y.}~\bibnamefont {Ni}}, \bibinfo {author} {\bibfnamefont
  {C.~A.}\ \bibnamefont {Pocs}}, \bibinfo {author} {\bibfnamefont
  {Y.}~\bibnamefont {Zhang}}, \bibinfo {author} {\bibfnamefont
  {F.}~\bibnamefont {Ye}}, \bibinfo {author} {\bibfnamefont {C.}~\bibnamefont
  {Hoffmann}}, \bibinfo {author} {\bibfnamefont {X.}~\bibnamefont {Wang}},
  \bibinfo {author} {\bibfnamefont {M.}~\bibnamefont {Lee}}, \bibinfo {author}
  {\bibfnamefont {M.}~\bibnamefont {Hermele}},\ and\ \bibinfo {author}
  {\bibfnamefont {I.}~\bibnamefont {Kimchi}},\ }\href
  {https://doi.org/10.1038/s41535-020-0232-6} {\bibfield  {journal} {\bibinfo
  {journal} {npj Quantum Materials}\ }\textbf {\bibinfo {volume} {5}},\
  \bibinfo {pages} {1} (\bibinfo {year} {2020})},\ \bibinfo {note} {publisher:
  Nature Publishing Group}\BibitemShut {NoStop}%
\bibitem [{\citenamefont {Ihmaïne}\ \emph {et~al.}(1988)\citenamefont
  {Ihmaïne}, \citenamefont {Perrin}, \citenamefont {Peña},\ and\
  \citenamefont {Sergent}}]{ihmaine_structure__1988}%
  \BibitemOpen
  \bibfield  {author} {\bibinfo {author} {\bibfnamefont {S.}~\bibnamefont
  {Ihmaïne}}, \bibinfo {author} {\bibfnamefont {C.}~\bibnamefont {Perrin}},
  \bibinfo {author} {\bibfnamefont {O.}~\bibnamefont {Peña}},\ and\ \bibinfo
  {author} {\bibfnamefont {M.}~\bibnamefont {Sergent}},\ }\href
  {https://doi.org/10.1016/0022-5088(88)90097-5} {\bibfield  {journal}
  {\bibinfo  {journal} {Journal of the Less Common Metals}\ }\textbf {\bibinfo
  {volume} {137}},\ \bibinfo {pages} {323} (\bibinfo {year}
  {1988})}\BibitemShut {NoStop}%
\bibitem [{\citenamefont {Neves}\ \emph {et~al.}(2024)\citenamefont {Neves},
  \citenamefont {Wakefield}, \citenamefont {Fang}, \citenamefont {Nguyen},
  \citenamefont {Ye},\ and\ \citenamefont {Checkelsky}}]{neves_crystal_2024}%
  \BibitemOpen
  \bibfield  {author} {\bibinfo {author} {\bibfnamefont {P.~M.}\ \bibnamefont
  {Neves}}, \bibinfo {author} {\bibfnamefont {J.~P.}\ \bibnamefont
  {Wakefield}}, \bibinfo {author} {\bibfnamefont {S.}~\bibnamefont {Fang}},
  \bibinfo {author} {\bibfnamefont {H.}~\bibnamefont {Nguyen}}, \bibinfo
  {author} {\bibfnamefont {L.}~\bibnamefont {Ye}},\ and\ \bibinfo {author}
  {\bibfnamefont {J.~G.}\ \bibnamefont {Checkelsky}},\ }\href
  {https://doi.org/10.1038/s41524-024-01220-x} {\bibfield  {journal} {\bibinfo
  {journal} {npj Computational Materials}\ }\textbf {\bibinfo {volume} {10}},\
  \bibinfo {pages} {39} (\bibinfo {year} {2024})},\ \bibinfo {note} {publisher:
  Nature Publishing Group}\BibitemShut {NoStop}%
\bibitem [{\citenamefont {Jain}\ \emph {et~al.}(2013)\citenamefont {Jain},
  \citenamefont {Ong}, \citenamefont {Hautier}, \citenamefont {Chen},
  \citenamefont {Richards}, \citenamefont {Dacek}, \citenamefont {Cholia},
  \citenamefont {Gunter}, \citenamefont {Skinner}, \citenamefont {Ceder},\ and\
  \citenamefont {Persson}}]{jain_commentary_2013}%
  \BibitemOpen
  \bibfield  {author} {\bibinfo {author} {\bibfnamefont {A.}~\bibnamefont
  {Jain}}, \bibinfo {author} {\bibfnamefont {S.~P.}\ \bibnamefont {Ong}},
  \bibinfo {author} {\bibfnamefont {G.}~\bibnamefont {Hautier}}, \bibinfo
  {author} {\bibfnamefont {W.}~\bibnamefont {Chen}}, \bibinfo {author}
  {\bibfnamefont {W.~D.}\ \bibnamefont {Richards}}, \bibinfo {author}
  {\bibfnamefont {S.}~\bibnamefont {Dacek}}, \bibinfo {author} {\bibfnamefont
  {S.}~\bibnamefont {Cholia}}, \bibinfo {author} {\bibfnamefont
  {D.}~\bibnamefont {Gunter}}, \bibinfo {author} {\bibfnamefont
  {D.}~\bibnamefont {Skinner}}, \bibinfo {author} {\bibfnamefont
  {G.}~\bibnamefont {Ceder}},\ and\ \bibinfo {author} {\bibfnamefont {K.~A.}\
  \bibnamefont {Persson}},\ }\href {https://doi.org/10.1063/1.4812323}
  {\bibfield  {journal} {\bibinfo  {journal} {APL Materials}\ }\textbf
  {\bibinfo {volume} {1}},\ \bibinfo {pages} {011002} (\bibinfo {year}
  {2013})}\BibitemShut {NoStop}%
\bibitem [{\citenamefont {Komleva}\ \emph {et~al.}(2020)\citenamefont
  {Komleva}, \citenamefont {Khomskii},\ and\ \citenamefont
  {Streltsov}}]{komleva_three-site_2020}%
  \BibitemOpen
  \bibfield  {author} {\bibinfo {author} {\bibfnamefont {E.~V.}\ \bibnamefont
  {Komleva}}, \bibinfo {author} {\bibfnamefont {D.~I.}\ \bibnamefont
  {Khomskii}},\ and\ \bibinfo {author} {\bibfnamefont {S.~V.}\ \bibnamefont
  {Streltsov}},\ }\href {https://doi.org/10.1103/PhysRevB.102.174448}
  {\bibfield  {journal} {\bibinfo  {journal} {Physical Review B}\ }\textbf
  {\bibinfo {volume} {102}},\ \bibinfo {pages} {174448} (\bibinfo {year}
  {2020})}\BibitemShut {NoStop}%
\bibitem [{\citenamefont {Skaggs}\ \emph {et~al.}(2024)\citenamefont {Skaggs},
  \citenamefont {Siegfried}, \citenamefont {Cho}, \citenamefont {Xin},
  \citenamefont {Garlea}, \citenamefont {Taddei}, \citenamefont {Bhandari},
  \citenamefont {Croft}, \citenamefont {Ghimire}, \citenamefont {Jang},\ and\
  \citenamefont {Tan}}]{skaggs_ba4rumn2o10_2024}%
  \BibitemOpen
  \bibfield  {author} {\bibinfo {author} {\bibfnamefont {C.~M.}\ \bibnamefont
  {Skaggs}}, \bibinfo {author} {\bibfnamefont {P.~E.}\ \bibnamefont
  {Siegfried}}, \bibinfo {author} {\bibfnamefont {J.~S.}\ \bibnamefont {Cho}},
  \bibinfo {author} {\bibfnamefont {Y.}~\bibnamefont {Xin}}, \bibinfo {author}
  {\bibfnamefont {V.~O.}\ \bibnamefont {Garlea}}, \bibinfo {author}
  {\bibfnamefont {K.~M.}\ \bibnamefont {Taddei}}, \bibinfo {author}
  {\bibfnamefont {H.}~\bibnamefont {Bhandari}}, \bibinfo {author}
  {\bibfnamefont {M.}~\bibnamefont {Croft}}, \bibinfo {author} {\bibfnamefont
  {N.~J.}\ \bibnamefont {Ghimire}}, \bibinfo {author} {\bibfnamefont {J.~I.}\
  \bibnamefont {Jang}},\ and\ \bibinfo {author} {\bibfnamefont
  {X.}~\bibnamefont {Tan}},\ }\href
  {https://doi.org/10.1021/acs.chemmater.4c00586} {\bibfield  {journal}
  {\bibinfo  {journal} {Chemistry of Materials}\ }\textbf {\bibinfo {volume}
  {36}},\ \bibinfo {pages} {6053} (\bibinfo {year} {2024})},\ \bibinfo {note}
  {publisher: American Chemical Society}\BibitemShut {NoStop}%
\bibitem [{\citenamefont {Paulose}\ \emph {et~al.}(2008)\citenamefont
  {Paulose}, \citenamefont {Mohapatra},\ and\ \citenamefont
  {Sampathkumaran}}]{paulose_spin-chain__2008}%
  \BibitemOpen
  \bibfield  {author} {\bibinfo {author} {\bibfnamefont {P.~L.}\ \bibnamefont
  {Paulose}}, \bibinfo {author} {\bibfnamefont {N.}~\bibnamefont {Mohapatra}},\
  and\ \bibinfo {author} {\bibfnamefont {E.~V.}\ \bibnamefont
  {Sampathkumaran}},\ }\href {https://doi.org/10.1103/PhysRevB.77.172403}
  {\bibfield  {journal} {\bibinfo  {journal} {Physical Review B}\ }\textbf
  {\bibinfo {volume} {77}},\ \bibinfo {pages} {172403} (\bibinfo {year}
  {2008})},\ \bibinfo {note} {publisher: American Physical Society}\BibitemShut
  {NoStop}%
\bibitem [{\citenamefont {Eyert}\ \emph {et~al.}(2008)\citenamefont {Eyert},
  \citenamefont {Schwingenschlögl}, \citenamefont {Hackenberger},
  \citenamefont {Kopp}, \citenamefont {Frésard},\ and\ \citenamefont
  {Eckern}}]{eyert_magnetic_2008}%
  \BibitemOpen
  \bibfield  {author} {\bibinfo {author} {\bibfnamefont {V.}~\bibnamefont
  {Eyert}}, \bibinfo {author} {\bibfnamefont {U.}~\bibnamefont
  {Schwingenschlögl}}, \bibinfo {author} {\bibfnamefont {C.}~\bibnamefont
  {Hackenberger}}, \bibinfo {author} {\bibfnamefont {T.}~\bibnamefont {Kopp}},
  \bibinfo {author} {\bibfnamefont {R.}~\bibnamefont {Frésard}},\ and\
  \bibinfo {author} {\bibfnamefont {U.}~\bibnamefont {Eckern}},\ }\href
  {https://doi.org/10.1016/j.progsolidstchem.2006.04.001} {\bibfield  {journal}
  {\bibinfo  {journal} {Progress in Solid State Chemistry}\ }\textbf {\bibinfo
  {volume} {36}},\ \bibinfo {pages} {156} (\bibinfo {year} {2008})}\BibitemShut
  {NoStop}%
\bibitem [{\citenamefont {Hardy}\ \emph {et~al.}(2003)\citenamefont {Hardy},
  \citenamefont {Lees}, \citenamefont {Maignan}, \citenamefont {Hébert},
  \citenamefont {Flahaut}, \citenamefont {Martin},\ and\ \citenamefont
  {Paul}}]{hardy_specific_2003}%
  \BibitemOpen
  \bibfield  {author} {\bibinfo {author} {\bibfnamefont {V.}~\bibnamefont
  {Hardy}}, \bibinfo {author} {\bibfnamefont {M.~R.}\ \bibnamefont {Lees}},
  \bibinfo {author} {\bibfnamefont {A.}~\bibnamefont {Maignan}}, \bibinfo
  {author} {\bibfnamefont {S.}~\bibnamefont {Hébert}}, \bibinfo {author}
  {\bibfnamefont {D.}~\bibnamefont {Flahaut}}, \bibinfo {author} {\bibfnamefont
  {C.}~\bibnamefont {Martin}},\ and\ \bibinfo {author} {\bibfnamefont {D.~M.}\
  \bibnamefont {Paul}},\ }\href {https://doi.org/10.1088/0953-8984/15/33/307}
  {\bibfield  {journal} {\bibinfo  {journal} {Journal of Physics: Condensed
  Matter}\ }\textbf {\bibinfo {volume} {15}},\ \bibinfo {pages} {5737}
  (\bibinfo {year} {2003})}\BibitemShut {NoStop}%
\bibitem [{\citenamefont {Richard}\ \emph {et~al.}(2017)\citenamefont
  {Richard}, \citenamefont {Benayad}, \citenamefont {Colin},\ and\
  \citenamefont {Martinet}}]{richard_charge_2017}%
  \BibitemOpen
  \bibfield  {author} {\bibinfo {author} {\bibfnamefont {J.}~\bibnamefont
  {Richard}}, \bibinfo {author} {\bibfnamefont {A.}~\bibnamefont {Benayad}},
  \bibinfo {author} {\bibfnamefont {J.-F.}\ \bibnamefont {Colin}},\ and\
  \bibinfo {author} {\bibfnamefont {S.}~\bibnamefont {Martinet}},\ }\href
  {https://doi.org/10.1021/acs.jpcc.7b03979} {\bibfield  {journal} {\bibinfo
  {journal} {The Journal of Physical Chemistry C}\ }\textbf {\bibinfo {volume}
  {121}},\ \bibinfo {pages} {17096} (\bibinfo {year} {2017})}\BibitemShut
  {NoStop}%
\bibitem [{\citenamefont {Halasyamani}\ and\ \citenamefont
  {Poeppelmeier}(1998)}]{ncsoxides}%
  \BibitemOpen
  \bibfield  {author} {\bibinfo {author} {\bibfnamefont {P.~S.}\ \bibnamefont
  {Halasyamani}}\ and\ \bibinfo {author} {\bibfnamefont {K.~R.}\ \bibnamefont
  {Poeppelmeier}},\ }\href {https://doi.org/10.1021/cm980140w} {\bibfield
  {journal} {\bibinfo  {journal} {Chemistry of Materials}\ }\textbf {\bibinfo
  {volume} {10}},\ \bibinfo {pages} {2753} (\bibinfo {year} {1998})},\ \Eprint
  {https://arxiv.org/abs/https://doi.org/10.1021/cm980140w}
  {https://doi.org/10.1021/cm980140w} \BibitemShut {NoStop}%
\end{thebibliography}%

\clearpage
\appendix
\setcounter{figure}{0}
\renewcommand{\thefigure}{S.\arabic{figure}}
\renewcommand{\theHfigure}{S.\arabic{figure}}
\setcounter{table}{0}
\renewcommand{\thetable}{S.\arabic{table}}
\renewcommand{\theHtable}{S.\arabic{table}}

\begin{center}
\section*{Supporting Information for "A Structural Map of Potential Correlated Electron Molecular Orbital Materials"}

Md. Rajbanul Akhond, Alexandru B. Georgescu\\
mdakhond@iu.edu, georgesc@iu.edu\\
Department of Chemistry, 800 East Kirkwood Avenue, Indiana University, Bloomington, Indiana 47405, United States\\
\end{center}
\FloatBarrier
\section{Cluster Motifs and associated MOs}
\begin{table}[h!]
\centering
\renewcommand{\arraystretch}{1.3}
\begin{tabular}{|c|l|}
\hline
\textbf{Cluster Size} & \textbf{Top Point Groups} \\
\hline
2 & $D_{\infty h}$, $C_{\infty v}$ \\
\hline
3 & $D_{\infty h}$, $C_{2v}$, $D_{3h}$, $C_{s}$, $C_{\infty v}$ \\
\hline
4 & $C_{s}$, $C_{2v}$, $D_{\infty h}$, $D_{2h}$, $C_{3v}$, $C_{2h}$, $C_{1}$, $C_{\infty v}$, $T_{d}$, $C_{2}$ \\
\hline
5 & $C_{s}$, $C_{2v}$, $D_{\infty h}$, $C_{1}$, $D_{2h}$, $D_{3h}$, $C_{2h}$, $C_{\infty v}$, $C_{2}$, $C_{4v}$ \\
\hline
6 & $C_{s}$, $D_{\infty h}$, $D_{3d}$, $C_{2v}$, $D_{2h}$, $C_{2h}$, $O_{h}$, $C_{1}$, $D_{3h}$, $C_{\infty v}$ \\
\hline
7 & $C_{s}$, $C_{1}$, $D_{\infty h}$, $C_{2v}$, $T_{d}$, $D_{3d}$, $C_{2h}$, $D_{2h}$, $C_{i}$, $D_{3h}$ \\
\hline
8 & $C_{1}$, $C_{s}$, $C_{i}$, $C_{2v}$, $C_{2h}$, $D_{2h}$, $D_{\infty h}$, $C_{\infty v}$, $S_{4}$, $C_{2}$ \\
\hline
\end{tabular}
\caption{Top point groups for different cluster sizes}
\label{pgt}
\end{table}

\begin{figure}[!ht]
\centering
\includegraphics[width=1\linewidth]{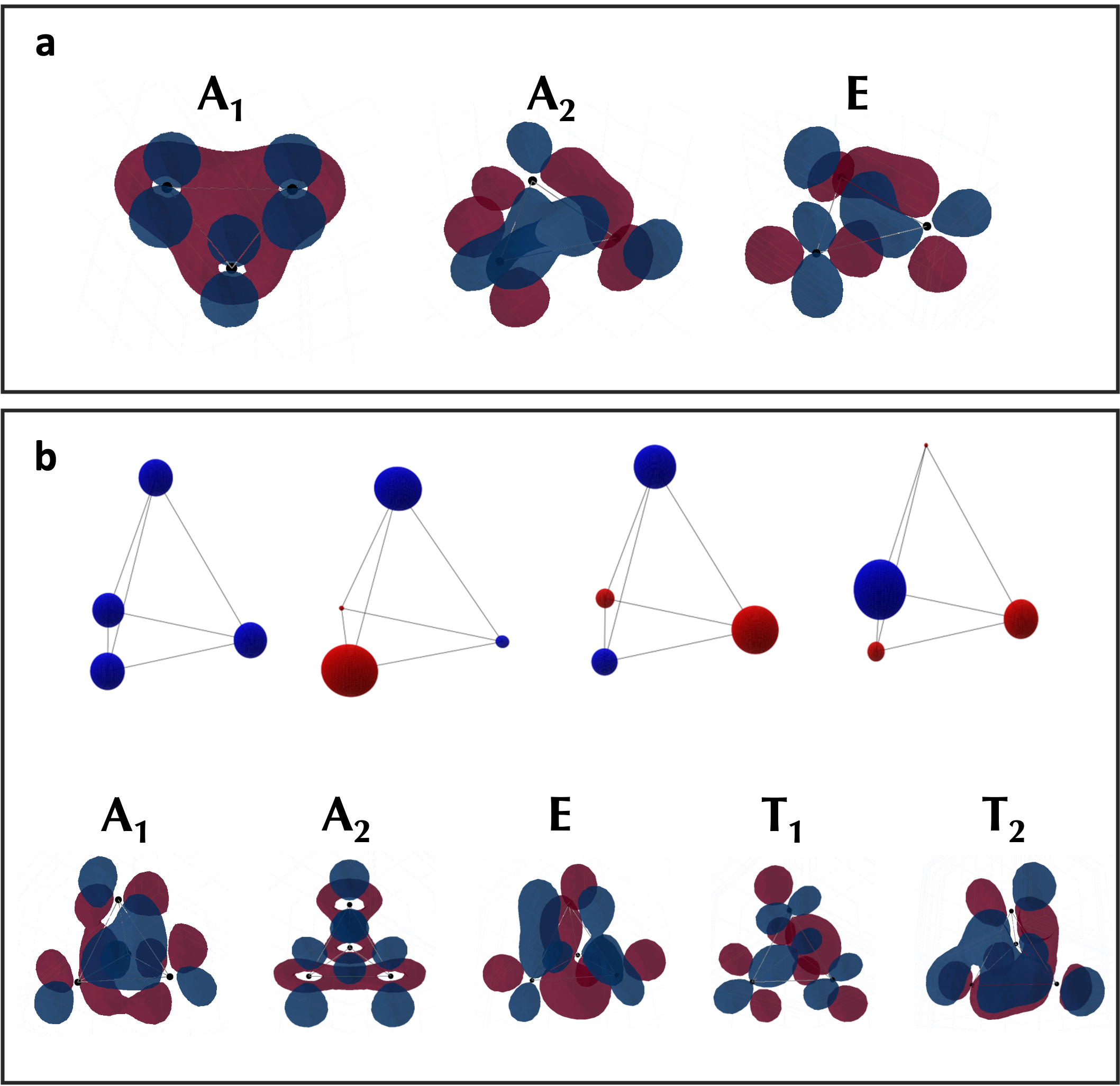}
\caption{(a) SALC basis orbitals for trimer cluster with $D_{3h}$ point group symmetry and (b) Tetrameter cluster motif with $T_d$ point group spanned with s-like isomorphic orbital basis and SALC(s) basis made with directional d-orbitals.}
\label{fig:salc}
\end{figure}

\clearpage
\section{Cluster Materials Explorer}
\begin{figure}[!ht]
  \centering
  \includegraphics[width=0.8\linewidth]{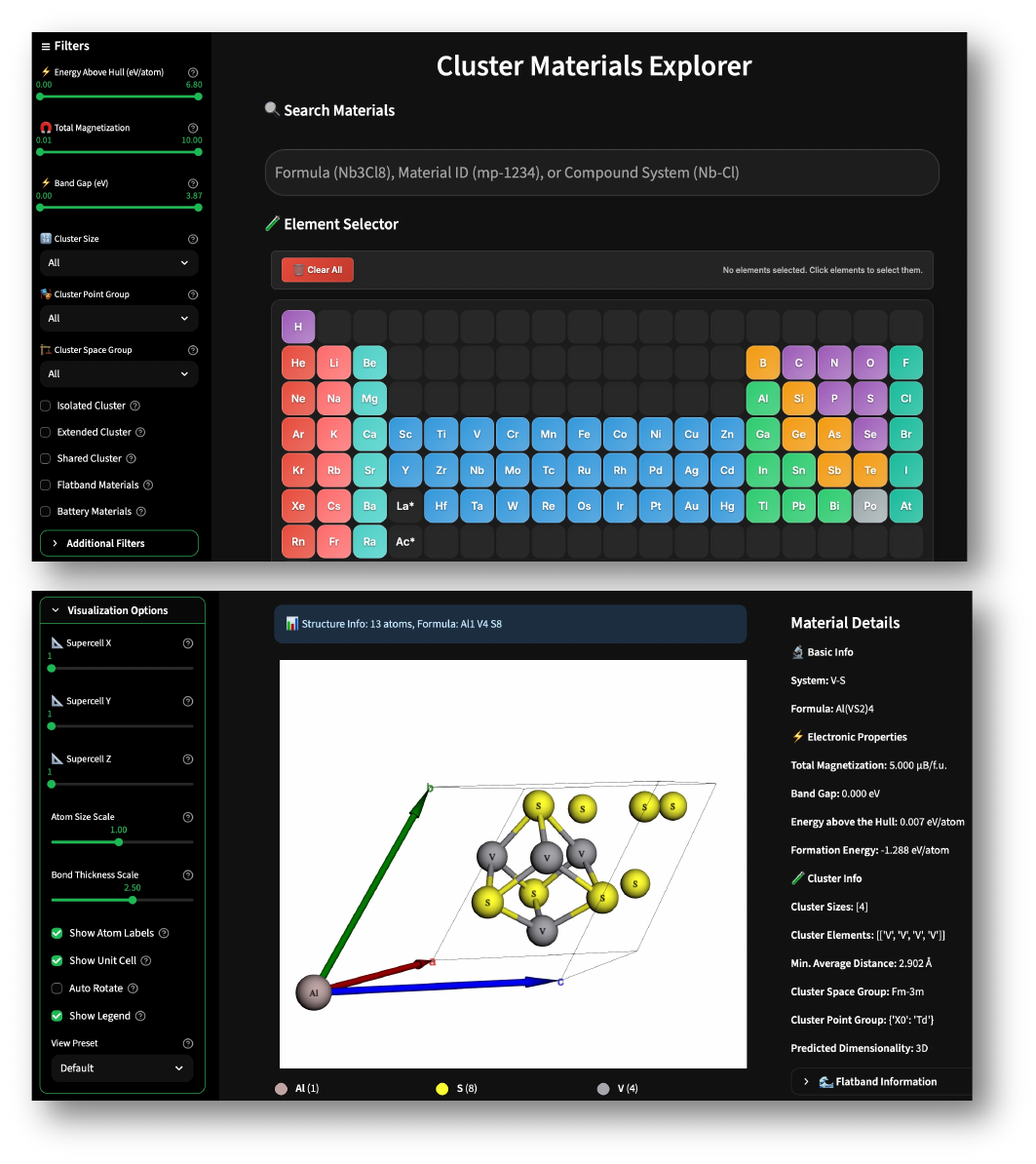}
  \caption{The "Cluster Materials Explorer" web-app where the dataset generated by Cluster Finder can be searched by formula, materials project id, elements, cluster type, and filtered by materials and cluster properties (Top). It allows for the visualization of selected materials and show details about specific materials, including flatband and battery properties (Bottom). }
  \label{fig:cme}
\end{figure}

\clearpage
\section{Cluster Motif Connectivity}
\begin{figure}[!ht]
\centering
\includegraphics[width=1\linewidth]{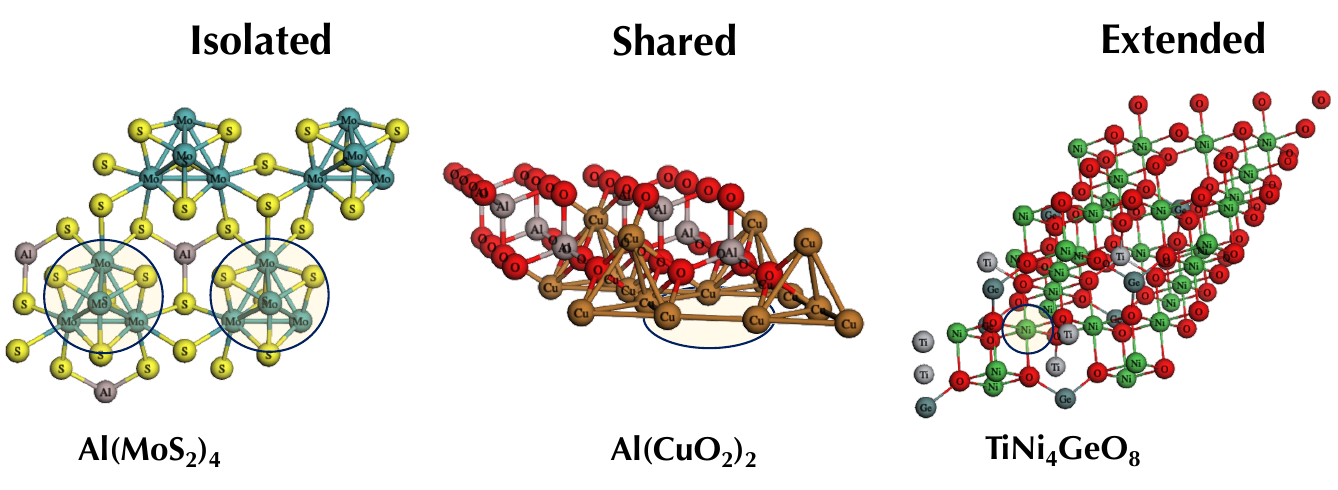}
\caption{Different types of cluster motif sharing. An example cluster of size four with $T_d$ cluster point group symmetry is shown, where \ce{Al(MoS2)4} contains isolated Mo cluster motifs, \ce{Al(CuO2)2} contains Cu cluster motifs with bond sharing and \ce{TiNi4GeO8} contains coalescence of Ni cluster motifs extending in 3d.}
\label{fig:connectivity}
\end{figure}

Due to the large number of materials that may display electronic properties arising from short metal-metal bonds in a solid, we also classified other type of materials - these may not all display CEMO properties as the clusters are no longer isolated. This can be seen in Figure \ref{fig:connectivity}. Beyond isolated clusters, we have identified extended cluster motifs, where the cluster motif in one cell is connected to the next cell. Another type of connectivity is shared, where as we make a supercell the cluster motif itself changes in terms of size and geometry.

To indentify cluster connectivity we can utilize the average cluster distance, however, in terms of extending the unit cell in this case. So, we build a $2*2*2$ supercell and compare how the cluster average distance chages compared to unit cell. If supercell has larger cluster with same average distance then it is flagged as extended cluster and if the average distance is different then it is flagged as shared cluster, otherwise the cluster is flagged as isolated and we go through the remaining steps as before. 

\clearpage
\section{Breathing Distortion and Cluster formation}
\begin{figure}[!ht]
\centering
\includegraphics[width=1\linewidth]{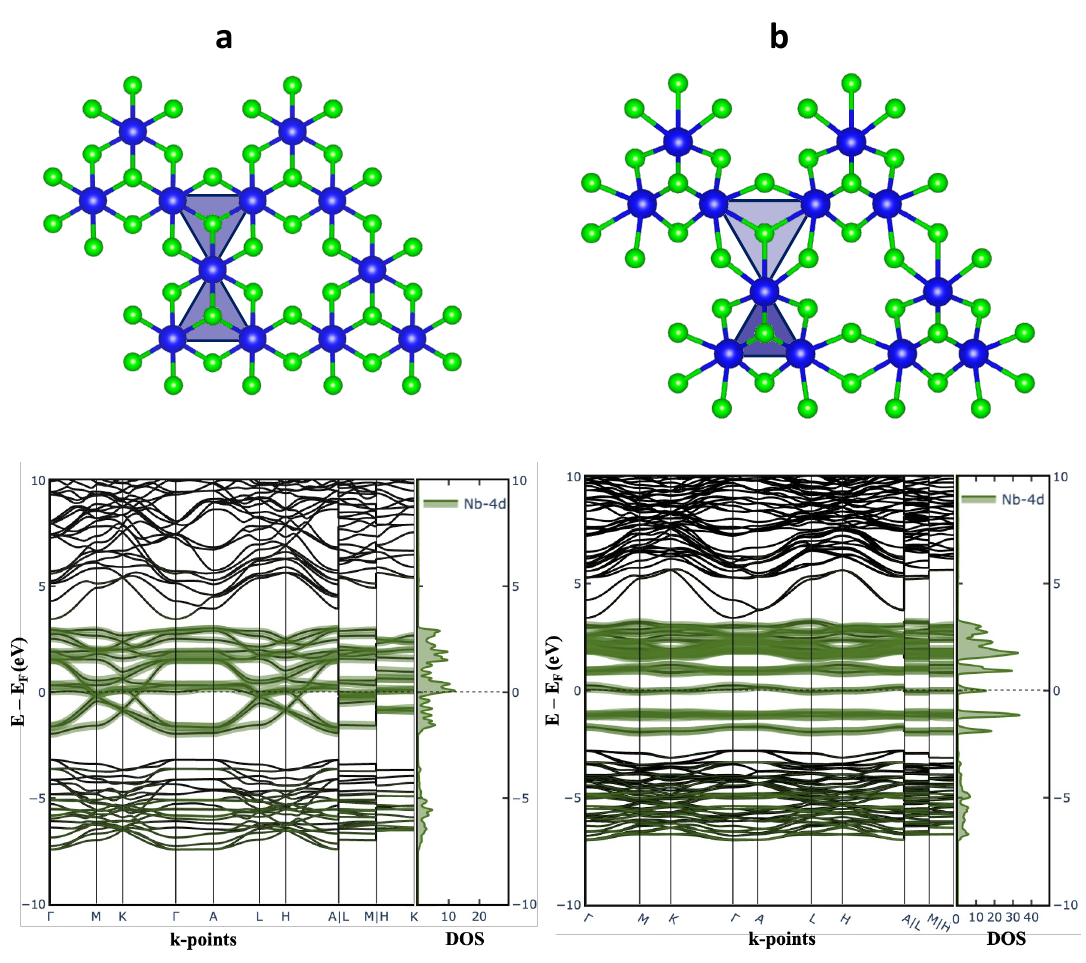}
\caption{Breathing distortion (BD) in kagome \ce{Nb3Cl8}. (a) Without BD, the Nb kagome network does not form trimers and no isolated flat band appears near the Fermi level. (b) With BD, the Nb atoms trimerize into \ce{Nb3} clusters, producing localized molecular orbitals and a flat band.}
\label{fig:bd}
\end{figure}

\begin{figure}[!ht]
\centering
\includegraphics[width=1\linewidth]{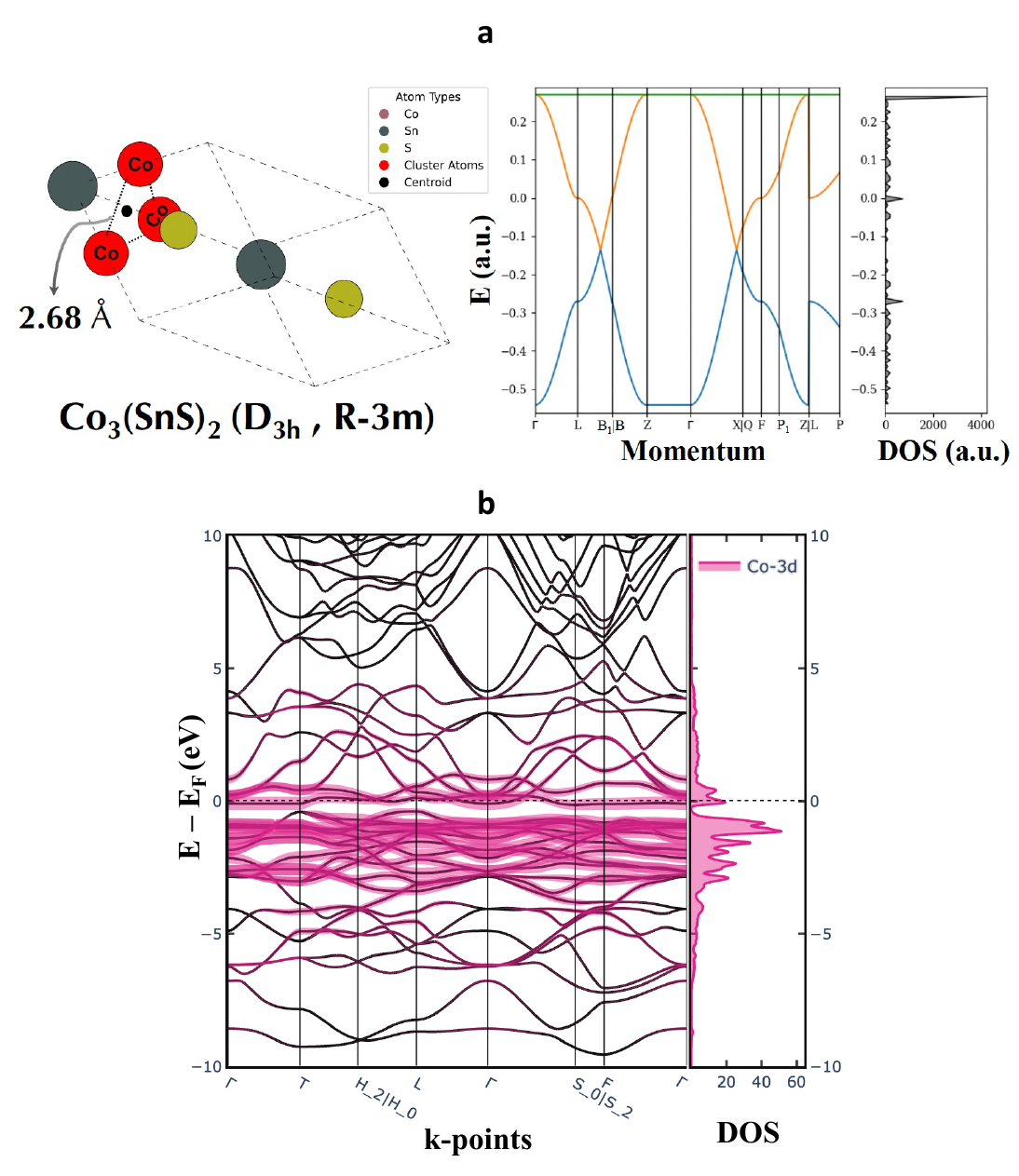}
\caption{Electronic structure of kagome \ce{Co3(SnS)2}, shown as a frustrated kagome example without breathing-distortion-driven trimerization. (a) Structural motif and tight-binding band picture. (b) Calculated band structure and projected density of states, showing that in the absence of cluster formation the kagome network does not produce isolated narrow molecular-orbital flat bands, in contrast to trimerized systems such as \ce{Nb3Cl8}.}
\label{fig:real}
\end{figure}

\clearpage
\section{Example CEMO Materials}
\begin{figure}[!ht]
\centering
\includegraphics[width=.9\linewidth]{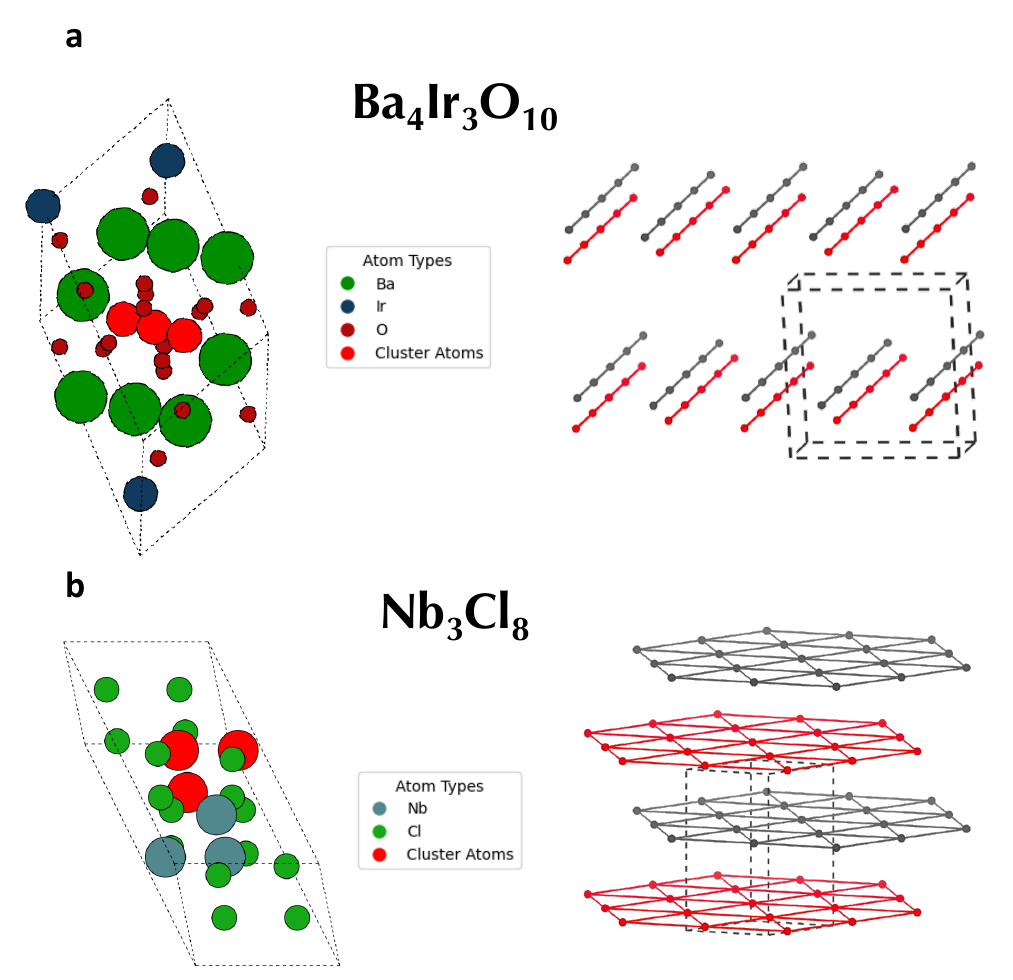}
\caption{Identified (a) linear trimer in \ce{Ba4Ir3O10} with its effective 1D dimensionality and (b) triangular trimer in \ce{Nb3Cl8} with its effective 2D dimensionality as predicted by our method and matches the \cite{cao_quantum_2020} and \cite{grytsiuk_nb3cl8_2024}, respectively.}
\label{fig:dim}

\end{figure}

\begin{figure}[!ht]
\centering
\includegraphics[width=1\linewidth]{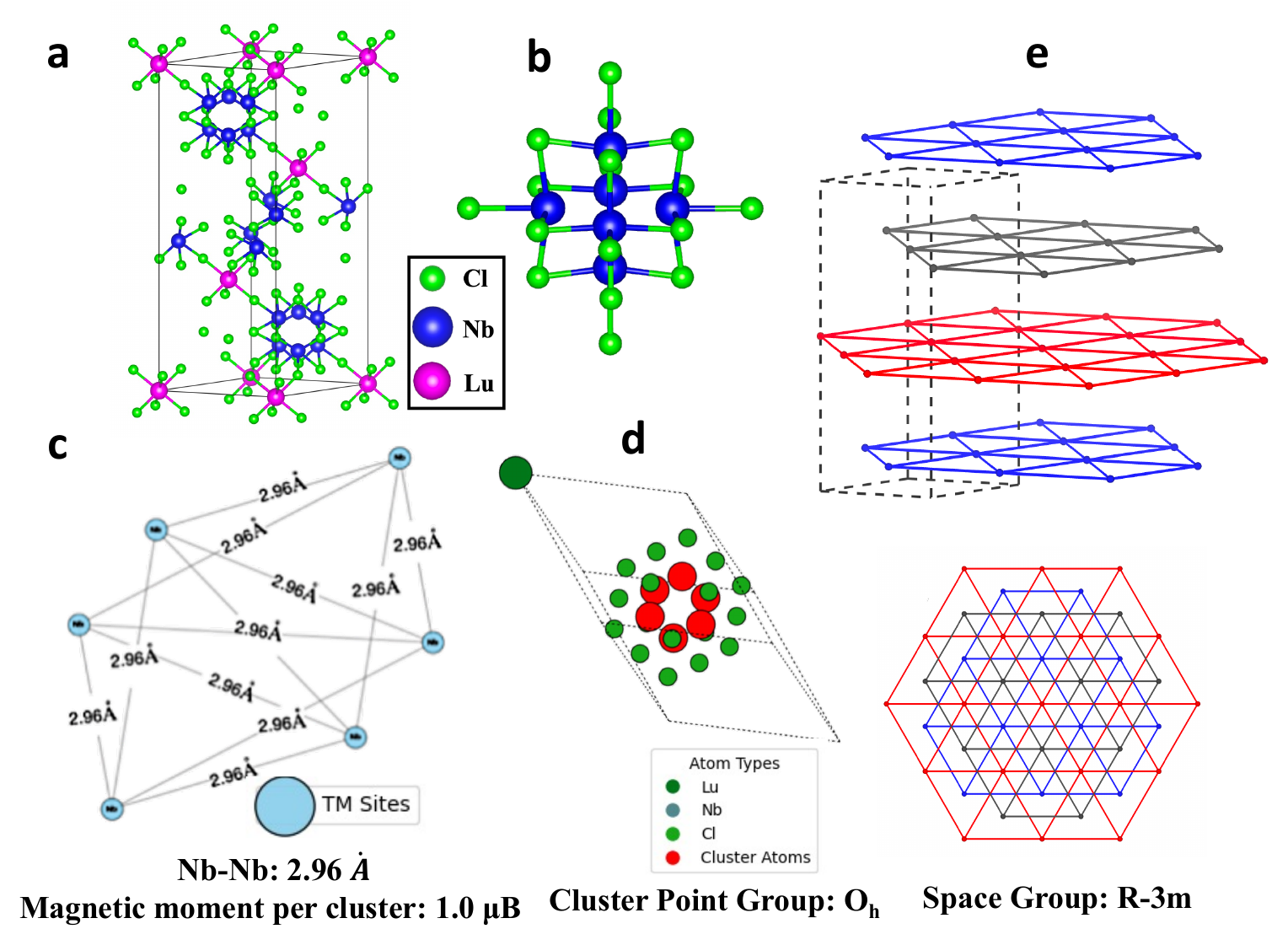}
\caption{Identification of a cluster material using our methods. (a) Crystal structure of \ce{Lu(NbCl3)6} as stored in the Materials Project database. (b) Visualization of a \ce{Nb6} cluster within the structure. (c) Graph representation of Nb connectivity (blue circles), highlighting short metal-metal distances. (d) Automated identification of cluster atoms \& point group within the structure. (e) Representation of the pseudo-2D Bravais lattice, where each point corresponds to a cluster, enabling symmetry analysis.}
\label{fig:ucluster}
\end{figure}

\clearpage
\section{Spin polarized Band structure of \ce{Fe16Rb28Te32}}
\begin{figure}[!ht]
\centering
\includegraphics[width=0.8\linewidth]{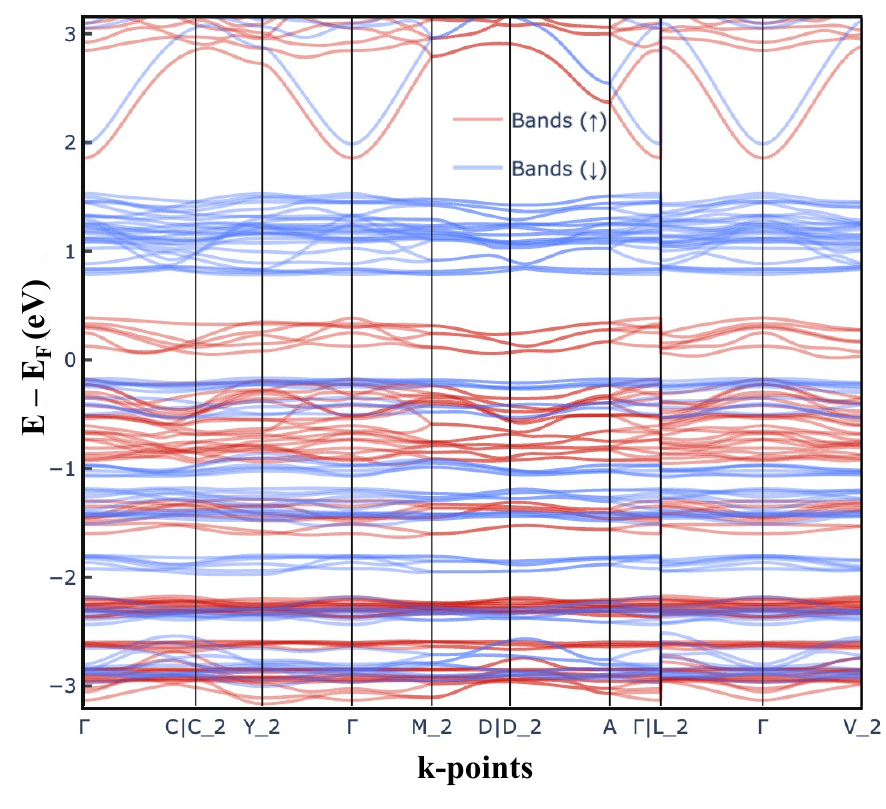}
\caption{Spin-polarized band structure calculation of \ce{Fe16Rb28Te32}.}
\label{fig:Fe16Rb28Te32_sp}
\end{figure}

\FloatBarrier
\section{Symmetry allowed materials properties}
The symmetry analysis extends to the full crystal lattice using the criteria shown in figure A1 and described in the work by P. Shiv Halasyamani and Kenneth R. Poeppelmeier, allowing classification into polar and nonpolar materials, with polar materials prioritized for further studies due to their potential for applications in multiferroics and ferroelectrics. Thermodynamic stability is assessed through proximity to the convex hull (within 0.1 eV/atom) using existing databases such as the Materials Project.
\begin{figure}[!ht]
\centering
\includegraphics[width=0.75\linewidth]{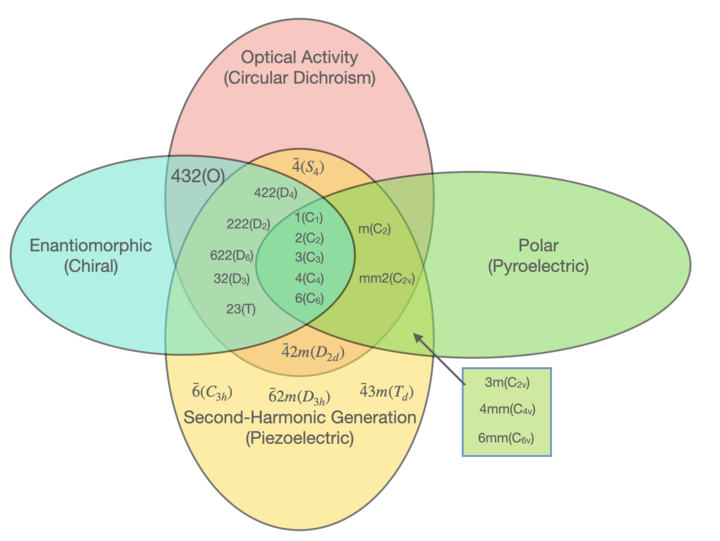}
\caption{Venn diagram of crystallographic point groups
classified by their symmetry-allowed physical properties,
including enantiomorphism (chirality), optical activity
(circular dichroism), polarity (pyroelectricity), and
piezoelectricity (e.g., second-harmonic generation). Point
groups are labeled using both Hermann–Mauguin and
Schoenflies notation. Adapted from Halasyamani and
Poeppelmeier \cite{ncsoxides}}
\label{fig:venn}
\end{figure}

\FloatBarrier
\section{Principal Component Analysis (PCA)}
PCA is a statistical technique that is used to reduce the dimensionality of a dataset while preserving as much variance as possible. In matrix terms, it involves transforming the original data into a new coordinate system defined by the principal components—directions in which the data varies the most.

Suppose we have $n$ data points in $d$-dimensional space (e.g., 3D with $(x, y, z)$ coordinates). This is represented as an $n \times d$ matrix $X$, where each row is a data point and each column is a dimension.
PCA requires centered data, so we compute the mean of each column (dimension), resulting in a $1 \times d$ mean vector $\mu$. Subtract $\mu$ from each row of $X$ to get the centered matrix:
\[
X_c = X - \frac{1}{n} \mathbf{1}_n \mu
\]
where $\mathbf{1}_n$ is an $n \times 1$ column vector of ones. The centered matrix $X_c$ remains $n \times d$.
Then, we compute the covariance matrix:
\[
C = \frac{1}{n-1} X_c^T X_c,
\]
which is $d \times d$. The eigenvectors of $C$ are the principal components, and the eigenvalues represent the variance along each component.
Diagonalizing $C$ matrix we get,
\[
C = V \Lambda V^T,
\]
where $V$ is a $d \times d$ matrix whose columns are the eigenvectors (principal directions), and $\Lambda$ is a $d \times d$ diagonal matrix of eigenvalues (variances). However, computing $C$ and its eigen decomposition can be inefficient for large $n$.
PCA can be computed more efficiently using Singular Value Decomposition (SVD) of the centered data matrix $X_c$, avoiding the explicit computation of $C$.
SVD is a factorization of any matrix into three matrices, providing insight into its structure and enabling applications like PCA.

For a matrix $A$ of size $m \times n$:
\[
A = U \Sigma V^T
\]
where, $U$: An $m \times m$ orthogonal matrix whose columns are the left singular vectors. $\Sigma$: An $m \times n$ rectangular diagonal matrix with non-negative singular values $\sigma_1 \geq \sigma_2 \geq \dots \geq \sigma_{\min(m,n)} \geq 0$ on the diagonal. $V$: An $n \times n$ orthogonal matrix whose columns are the right singular vectors (principal directions in PCA). $V^T$: The transpose of $V$, size $n \times n$.
For a reduced SVD calculation the decomposition is “economy-sized”, where,  $U$ becomes $m \times \min(m,n)$,  $\Sigma$ is a vector of length $\min(m,n)$ (singular values) and $V^T$ is $\min(m,n) \times n$.
To calculate PCA for $X_c$ (size $n \times d$) we apply SVD as, $X_c = U \Sigma V^T$. Where, $V$ (from $V^T$) gives the principal directions (columns are eigenvectors of $C$) and  Singular values $\sigma_i$ in $\Sigma$ relate to the eigenvalues of $C$ as,
\[
\lambda_i = \frac{\sigma_i^2}{n-1},
\]
where $\lambda_i$ is the variance along the $i$-th principal component.
The number of significant singular values indicates the effective dimensionality of the data.
\end{document}